\documentclass[journal,12pt,onecolumn,draftclsnofoot]{IEEEtran}
\usepackage[T1]{fontenc}
\usepackage{times}
\usepackage{color}
\usepackage{amsmath}
\usepackage{amssymb}
\usepackage{verbatim}
\usepackage{stmaryrd}
\usepackage{stackrel}
\usepackage{graphicx}
\usepackage{cite}
\usepackage{cases}
\usepackage{caption}
\usepackage{stackrel}
\usepackage{multicol}
\usepackage[USenglish]{babel}
\usepackage{array}
\usepackage{multirow}
\usepackage{amsfonts}
\usepackage{float}
\usepackage{balance}
\usepackage{ragged2e}
\usepackage{subfigure}
\usepackage{physics}
\usepackage[ruled,linesnumbered,noend]{algorithm2e}
\usepackage{setspace}
\usepackage{stfloats}
\usepackage{tikz}
\usepackage{booktabs}
\usepackage{autobreak}
\usepackage{enumerate}
\usepackage{comment}

\renewcommand{\raggedright}{\leftskip=0pt \rightskip=0pt plus 0cm}
\raggedright

\allowdisplaybreaks
\SetAlgoSkip{bigskip}

\begin{document}
\title{Radio Resource Management for Cellular-Connected UAV: A Learning Approach}
\author{Yuanjian~Li,~\IEEEmembership{Graduate Student Member,~IEEE} and A. Hamid~Aghvami,~\IEEEmembership{Life Fellow,~IEEE}\thanks{Yuanjian Li and A. Hamid Aghvami are with Centre for Telecommunications Research (CTR), King's College London, London WC2R 2LS, U.K. (e-mail: \{yuanjian.li, hamid.aghvami\}@kcl.ac.uk)}
\thanks{This work has been submitted to the IEEE for possible publication.  Copyright may be transferred without notice, after which this version may no longer be accessible.}
}
\maketitle
\begin{abstract}
Integrating unmanned aerial vehicles (UAVs) into existing cellular networks encounters lots of challenges, {among which one of the most striking concerns is how to achieve harmonious coexistence of aerial transceivers, inter alia, UAVs, and terrestrial user equipments (UEs).} In this paper, a cellular-connected UAV network is focused, where multiple UAVs receive messages from base stations (BSs) in the down-link, while BSs are serving ground UEs in their cells. {For effectively managing inter-cell interferences (ICIs) among UEs due to intense reuse of time-frequency resource block (RB) resource, a first $p$-tier based RB coordination criterion is proposed and adopted. Then, to enhance wireless transmission quality for UAVs while protecting terrestrial UEs from being interfered by ground-to-air (G2A) transmissions, a radio resource management (RRM) problem of joint dynamic RB coordination and time-varying beamforming design minimizing UAV's ergodic outage duration (EOD) is investigated. To cope with conventional optimization techniques' inefficiency in solving the formulated RRM problem,} a deep reinforcement learning (DRL)-aided solution is initiated, where deep double duelling Q network (D3QN) and twin delayed deep deterministic policy gradient (TD3) are invoked to deal with RB coordination in discrete action domain and beamforming design in continuous action regime, respectively. {The hybrid D3QN-TD3 solution is trained via interacting with the considered outer and inner environments in an online centralized manner so that it can then help achieve the suboptimal EOD minimization performance during its offline decentralized exploitation phase. Simulation results have illustrated the effectiveness of the proposed hybrid D3QN-TD3 algorithm, compared to several representative baselines.}
\end{abstract}

\section{Introduction}
{In current markets, unmanned aerial vehicles (UAVs) \cite{alsamhi2021green,wang2022uav,li2021energy} are commonly communicating with their ground-based pilots via simple point-to-point (P2P) links over unlicensed spectrum, e.g., the industrial, scientific and medical (ISM) band at 2.4 GHz, leading to inferior ground-to-air (G2A) transmission performance, including low data throughput, limited communication range and interference vulnerability \cite{liu2019multi}. For realizing UAVs' large-scale deployment and further improving G2A communication quality, one promising approach is to integrate UAVs into worldwide-deployed cellular networks as aerial user equipment (UE), leveraging powerful terrestrial BSs to support UAVs, widely known as cellular-connected UAV solution \cite{zeng2019path,zeng2021simultaneous,li2022path,zhang2018cellular,wu2021comprehensive,chandhar2017massive,senadhira2020uplink,mei2019cellular,pang2019uplink,mei2019cooperative}. In contrast to P2P aerial-terrestrial communications, cellular-connected UAV technique can help establish beyond-visual-and-radio-line-of-sight (BVRLoS) communications between terrestrial BSs and UAVs, which is beneficial for realizing long-distance UAV application immune to range limitation, not to mention other advantages such as enhanced performance of reliability, security, transmission rate and coverage \cite{zhang2018cellular,chandhar2017massive,wu2021comprehensive}. Besides, cellular-connected UAV is cost-effective because countless cellular BSs worldwide can be reused to support A2G communications, with no requirement on dedicated infrastructure reconstruction \cite{wu2021comprehensive}. %
} Unfortunately, the current cellular networks are exclusively established for serving ground UEs (GUEs), barely considering aerial UEs. Specifically, BSs' antennas in cellular networks are conventionally down-tilted towards the earth for mitigating inter-cell interferences (ICIs), which means that UAVs can only be served via the side-lobes and thus satisfactory G2A connections cannot be guaranteed in general \cite{hattab2020energy,li2022intelligent}. From the perspective of forthcoming 5G-beyond (B5G) or 6G cellular networks, the main serving objects are still GUEs, raising that finding a proper way of involving UAVs into cellular networks without posing negative impacts on terrestrial transmissions is inherently of importance. 
{\subsection{Related Works}}
Unlike terrestrial cellular transmissions where non-line-of-sight (NLoS) pathloss appears more frequently, the first significant difference introduced by drones is that line-of-sight (LoS) link occurs more likely in G2A communications \cite{zhou2018uav,zhou2018mobile,chu2019uav,hu2019optimal,wang2020energy}, which plays the role as a double-edged blade. On one hand, LoS-dominant G2A links can help relieve the sufferance of severe multi-path fading, shadowing and pathloss, which are common illnesses in terrestrial transmissions due to vast existence of blockages, e.g., buildings and trees. On the other hand, it may make drones generate stronger interferences (or suffer more severe interferences) to (or from) BSs in the up-link (or the down-link) transmissions. Besides, drones can cover larger region for data transmissions {blessed by} their high flying altitudes, then greater \textit{macro-diversity gain} can usually be achieved because more BSs can cooperate to enhance G2A communication qualities in terms of throughput and reliability \cite{mei2019cooperative}. Unfortunately, more co-channel interfering sources for the drones in the down-link might be involved as well (or, UAVs can act as the interferers to more GUEs in the up-link) \cite{mei2019cellular}. Therefore, interference coordination issue for cellular-connected UAV networks is more intricate and must be seriously treated. Various interference management strategies have been investigated in the literature for terrestrial cellular transmission scenario, e.g., inter-cell interference coordination (ICIC) \cite{boudreau2009interference,kosta2012interference}, cognitive beamforming \cite{zhang2010dynamic} and coordinated multipoint (CoMP) communications \cite{irmer2011coordinated}. However, they are most likely ineffective to handle more sophisticated interfering environment caused by UAVs with LoS-dominant G2A links and {broader wireless} coverage. Therefore, interference management {strategies} that are adaptive to cellular-connected UAV networks should be delicately designed to achieve efficient spectrum sharing with coexisting GUEs. Up to date, there exist several related works devoted to offering interference management {solutions} for cellular-connected UAV networks \cite{chandhar2017massive,senadhira2020uplink,liu2019multi,mei2019cellular,pang2019uplink,mei2019cooperative}. Chandhar \textit{et al.} \cite{chandhar2017massive} leveraged multiple-input multiple-output (MIMO) technique to deal with interference coordination problem of single-antenna UAV swarms served by a multiple-antenna BS. Senadhira \textit{et al.} \cite{senadhira2020uplink} studied the impacts of UAV's trajectory and altitude for up-link non-orthogonal multiple access (NOMA) cellular-connected UAV network, in which ICI issue was dealt with NOMA technique. However, protecting the GUEs located in current cell or other cells within the coverage of UAVs was not considered in these works, which may significantly deteriorate the transmission performance of potential co-channel GUEs. Fortunately, some recent literature took care of interference control issue while protecting GUEs in cellular-connected UAV networks \cite{liu2019multi,mei2019cellular,pang2019uplink,mei2019cooperative}. Liu \textit{et al.} \cite{liu2019multi} proposed a new cooperative interference cancellation strategy for multi-beam cellular-connected UAV up-link transmissions, in which co-channel interference elimination and sum-rate maximization were investigated with the help of transmit beamforming design. Mei \textit{et al.} \cite{mei2019cellular} studied interference mitigation issue in up-link communications from a UAV to BSs, where weighted sum-rate of the UAV and GUEs was maximized via jointly optimizing up-link cell association and power allocation. {Pang \textit{et al.} \cite{pang2019uplink} investigated up-link transmission optimization problem on sum-rate of UAV and co-channel GUEs within NOMA-aided cellular-connected UAV networks, where successive interference cancellation (SIC) was implemented at both UAV-paired BSs and other BSs serving GUEs with the same band to cope with strong interferences generated by UAV to GUEs. Mei \textit{et al.} \cite{mei2019cooperative} proposed a BS-based cooperative beamforming (CB) scheme for cellular-connected UAV transmissions to suppress interferences caused by co-channel GUEs to UAV, where the formulated UAV's received signal-to-interference-plus-noise ratio (SINR) maximization problem was solved by a divide-and-conquer approach.} However, these works contain practical limitations. First, they assumed fixed-location UAV, without involving UAV's mobility. Second, the G2A channel models they applied are based on either oversimplified free-space pathloss channel model or slightly advanced probabilistic LoS channel model. It is worth noting that probabilistic G2A channel model can only reflect G2A pathloss gain in an expected manner without considering local building distribution where UAVs are actually deployed \cite{zeng2021simultaneous}. Last but not least, {traditional optimization-aided solutions they proposed are commonly inefficient in solving complex interference coordination problems due to non-convexity, even given strong assumptions of needed evaluation factors' availability.} 

{\subsection{Contributions}}
{Motivated by the above observations, radio resource management (RRM) issue of suboptimalbeamforming design within down-link cellular-connected UAV networks is considered in this paper, where the fundamental challenge of integrating UAVs into worldwide cellular networks that are designed delicately for serving GUEs is taken care of, while machine learning (ML)-native solution for achieving harmonious coexistence of non-terrestrial transceivers, i.e., UAVs, and terrestrial nodes, i.e., GUEs, is designed.} 

The main contributions of this paper can be summarised as follows.
\begin{itemize}
	\item {Different from related literature adopting statistical G2A channel model for mathematical tractability, e.g., probabilistic G2A channel model, LoS/NLoS G2A pathloss is tracked via checking potential blockages between UAV and BSs in this paper,} according to one realization of local building distribution suggested by International Telecommunication Union (ITU) \cite{series2013propagation}. The considered G2A channel model is more practical than its statistical counterpart reflecting ergodic pathloss gain over a large number of building distribution realizations {because the layout of local area can barely vary in real life.}
	\item A joint time-frequency resource block (RB) allocation and beamforming design optimization problem is formulated to minimize the ergodic outage duration (EOD) of {cellular-connected UAVs}, for arbitrary trajectory {and small-scale fading modelling.} Specifically, the RB allocation is utilized to assign proper RB resource to UAVs while {ensuring} that the terrestrial transmissions are not violated by the potential co-channel interferences generated from BSs {appointed} to serve UAVs. {To enhance G2A transmission quality after RB coordination,} transmit beamforming is invoked in the presence of imperfect G2A channel estimation. 
	\item {The practical consideration of building distribution based pathloss model and the generality of the formulated EOD minimization problem to trajectory and small-scale fading make it extremely difficult to be solved by classical optimization techniques, e.g., convex optimization. To cope with this hassle,} a deep reinforcement learning (DRL) \cite{wang2021jamming,ding20203d} solution is proposed {after mapping cellular environment into outer and inner Markov decision processes (MDPs), reflecting BSs' dynamic RB possessions and small-scale (fast) fading's time-varying characteristics, respectively.} The outer MDP contains discrete action space (i.e., RB indices), which is tackled by invoking deep double duelling Q network (D3QN), while the continuous action space (i.e., beamforming vectors) in the inner MDP is {coped} with twin delayed deep deterministic policy gradient (TD3) approach. {The hybrid D3QN-TD3 solution learns to optimize EOD performance via interacting with environments in the online centralized training phase, after which the trained D3QN and TD3 agents can be deployed to offer independent EOD performance gains in the phase of offline decentralized exploitation.}
\end{itemize}

{This paper contains plenty of notations and abbreviations, which may lead to difficulty in following, for relieving which Table \ref{table_notation_abbreviation} is provided. Please note that a more comprehensive list of notations and their descriptions can be found in Section \ref{Sec_numerical_results}, i.e., Table \ref{Simulation_Settings}.}

\begin{table} 
	\centering
	\scriptsize
	\belowcaptionskip=-.1cm\abovecaptionskip=0cm
	\caption{Descriptions of selected abbreviations and key notations} 
	\label{table_notation_abbreviation}
	\begin{tabular}{p{1.8cm}|p{5.5cm}||p{1.35cm}|p{4.8cm}} 
		\toprule
		\textbf{Abbreviations} & \textbf{Description}  & \textbf{Notations} & \textbf{Description}\\ 
		\midrule
		BS/G2A/UE& base station/ground-to-air/user equipment & $b/u/g/k$ & notation for denoting BS/DUE/GUE/RB\\
		GUE/DUE/ICI & ground UE/drone UE/inter-cell interference & $\mathcal{B}/\mathcal{U}/\mathcal{G}/\mathcal{K}$ & set of BSs/DUEs/GUEs/RBs\\
		LoS/NLoS/RB & line-of-sight/non-LoS/resource block & $\vec{q}_{b}/\vec{q}_{u}/\vec{q}_{g}$ & coordinate vector of BS/DUE/GUE\\
		EOD/RBP & ergodic outage duration/RB possession &$\mathcal{TI}_{b}(p)$ & set of first $p$-tier BSs encompassing BS $b$\\            
		D3QN & deep double duelling Q network & $\mathcal{B}_{o}^{k}/\hat{\mathcal{B}}_{o}^{k}/\breve{\mathcal{B}}_{o}^{k}$ & set of occupied/potential/available BSs\\            
		TD3 & twin delayed deep deterministic policy gradient & $card(\cdot)/C_{u}^{k}$ & set's cardinality/RB association indicator\\             
		ITU& international telecommunication union &$\vec{h}/\vec{w}$ & small-scale fading/beamforming vector\\              
		B2G/{B2D} & BS to GUE/DUE &$\mathbb{E}/\mathcal{C}$ & mathematical expectation/RBP map\\       
		CNN/RBP & convolutional neural network/RB Possession &$\dagger/\nabla$ & Hermitian transpose/taking gradient\\        
		RSRP/RSRQ & reference signal received power/quality &$\preceq/\Vert\cdot\Vert$ & element-wise inequality/Euclidean norm\\     
		\bottomrule
	\end{tabular}
\end{table}

\section{System Model and Problem Formulation}
\label{d3qn_ddpg_section_system_model}

\begin{figure}[htbp]
	\centering  
	\subfigcapskip=-.1cm
	\subfigure[System model]{
		\label{System Model}
	\includegraphics[width=.44\linewidth]{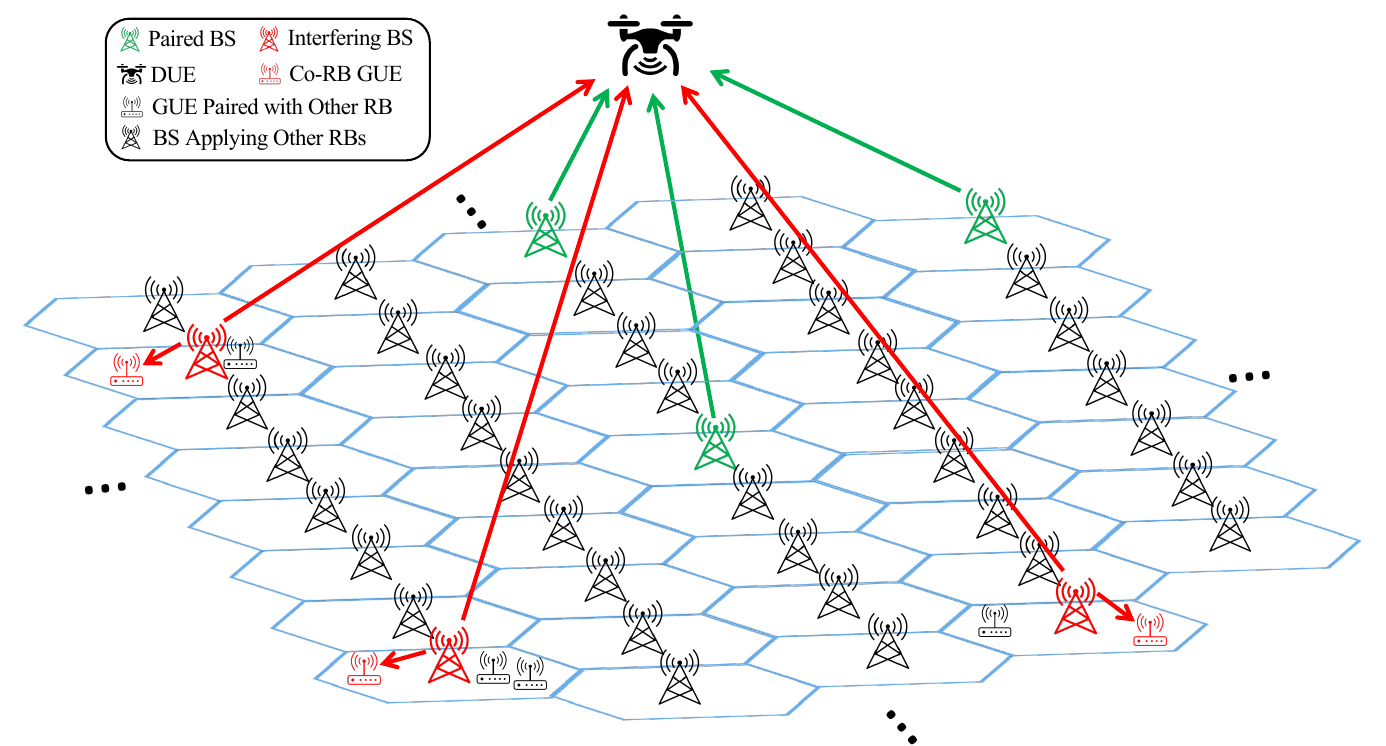}}
	\subfigure[First $p$-tier BSs]{
		\label{RB_Diagram_Tier3}
		\includegraphics[width=.245\linewidth]{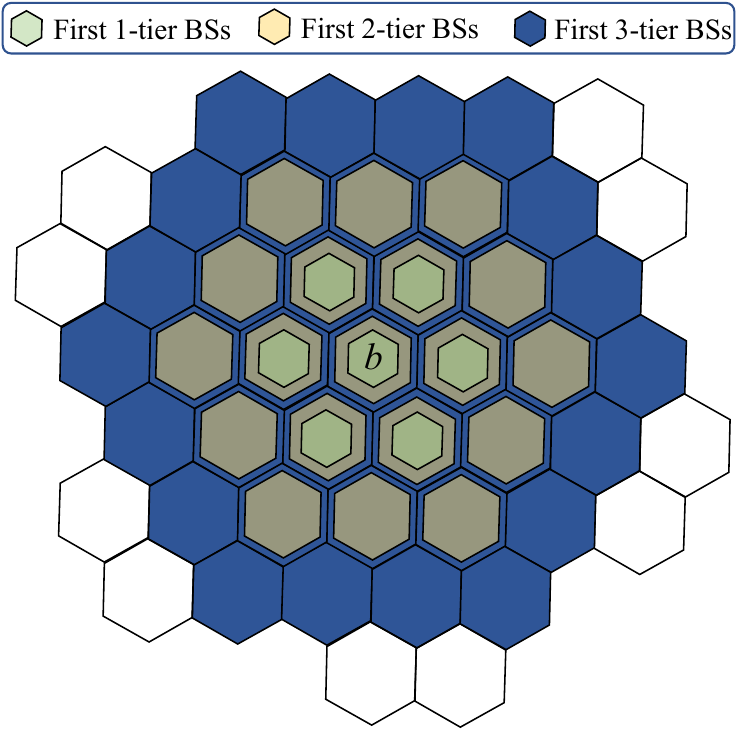}}\hspace{.3cm}
	\subfigure[An example of BS grouping]{
		\label{RB_Allocation_Example}
		\includegraphics[width=.24\linewidth]{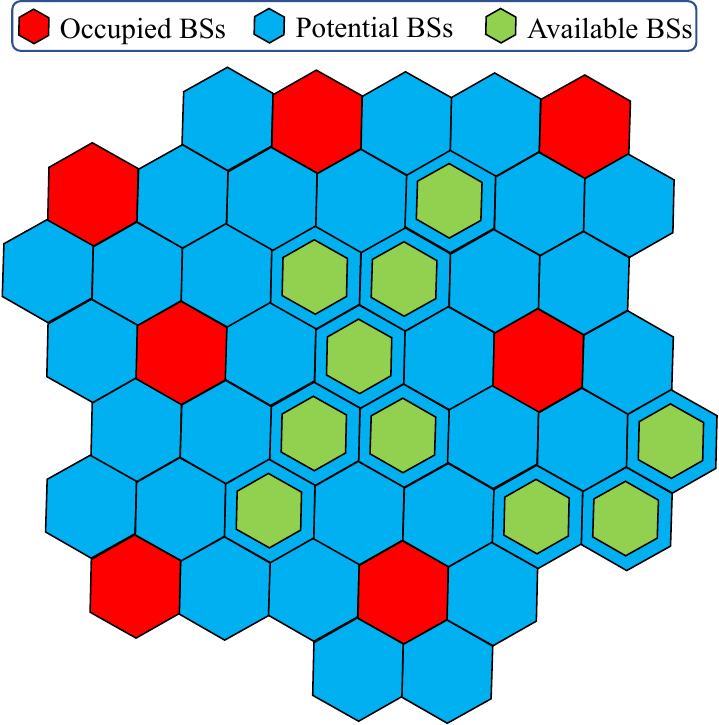}}
	\caption{Illustrations of system model, the first $p$-tier set of BS $b$ and an instance of BS grouping in the case of $p=1$, where $card(\mathcal{TI}_{b}(1))=7$, $card(\mathcal{TI}_{b}(2))=19$ and $card(\mathcal{TI}_{b}(3))=37$}
	\label{RB_Tier_and_Example}
\end{figure}

In this paper, RRM problem of RB allocation and beamforming design for down-link cellular-connected UAV network is considered, where a set $\mathcal{B}=\{1,\dots,B\}$ of $B$ terrestrial BSs serves a set $\mathcal{U}=\{1,\dots,U\}$ of $U$ drone UEs (DUEs) and a set $\mathcal{G}=\{1,\dots,G\}$ of $G$ GUEs using a set $\mathcal{K}=\{1,\dots,K\}$ of $K$ RBs at each BS, within a given subregion of cellular network, e.g., Fig. \ref{System Model}. Each DUE is assumed to equip single antenna for receiving wireless information and so as each GUE, while all the terrestrial BSs employ antenna array for message emitting. Specifically, each terrestrial BS $b\in\mathcal{B}$ possesses $M$ antennas, serving $g_{b}$ GUEs with orthogonal RBs (so there does not exist intra-cell interferences within each cell), where $g_{b}\geq1, \forall b\in\mathcal{B}$ and $\sum_{b=1}^{B}g_{b}=G$. Different from terrestrial transmission scenario, DUEs fly in the sky {with} relatively high altitudes, resulting in higher probability of achieving LoS-dominant {G2A} links from BSs. Thus, DUEs are able to connect with more BSs within their wireless coverage.
However, this characteristic may not only induce more and stronger desired signals but also result in richer co-channel interferences. To practically reflect the aforementioned double-edged blade feature, each DUE is considered to be associated with at least one BS when possible, taking advantage of macro-diversity gain from terrestrial BSs. 
Unfortunately, the assigned RB for a DUE might be already occupied by some GUEs due to heavy frequency reuse in cellular networks, severely interfering DUE via LoS-dominant channels. Therefore, RB allocation plays an important role in the considered cellular-connected UAV network. Besides, after RB assignment for a DUE, wireless transmission performance can be enhanced via invoking transmit beamforming technique at the corresponding serving BSs. Note that we do not consider transmit power control strategy at each BS, and thus we fix $P_{b}=P$ for all terrestrial BSs.\footnote{Transmit power control is indeed an important approach for interference management in cellular networks. In our considered model, it is straightforward to infer that all BSs should communicate with their paired DUEs using maximum transmit power, {which may cause stronger ICIs to co-channel GUEs.} Besides, all the occupied BSs are supposed to apply their minimum transmit power to reduce the level of co-channel interference to DUEs, which inevitably deteriorates the transmission quality for their severing GUEs. Therefore, to tackle this dilemma, we fix the transmit powers of all considered BSs as a constant $P$.}

The 3-dimensional (3D) locations of each DUE, each ground BS and each GUE are denoted as $\vec{q}_{u}=(x_{u},y_{u},h_{u})$, $\vec{q}_{b}=(x_{b},y_{b},z_{b})$ and $\vec{q}_{g}=(x_{g},y_{g},0)$, respectively. For simplicity and without loss of generality, the flying altitude of each DUE is assumed universally as $h_{u}=h$ and the height of each BS's antenna is set identically as $z_{b}=z$, where $h\gg z$ stands. Each DUE is supposed to reach its destination $\vec{q}_{u}(D)$ from predefined initial location $\vec{q}_{u}(I)$ with time cost $T_{u}$.\footnote{The application of UAVs as DUEs corresponds to various use cases, inter alia, cargo delivery and aerial inspection, where DUEs have to maintain solid connectivity with terrestrial BSs for accomplishing their respective missions \cite{zeng2021simultaneous}, especially for enabling BVRLoS piloting \cite{wu2021comprehensive}. For specific DUE $u\in\mathcal{U}$, the flying duration $T_{u}$ is determined by its trajectory and velocity.} 

For clarity, the considered subregion is formulated as a cubic sphere specified by $[x_{\text{lo}}, x_{\text{up}}]\times[y_{\text{lo}}, y_{\text{up}}]\times[z_{\text{lo}}, z_{\text{up}}]$, where the subscripts "lo" and "up" represent the lower and upper boarders of this 3D airspace, respectively. Furthermore, the coordinate of arbitrary DUE $u$ at time $t\in[0, T_{u}]$ should {locate} in the range of $\vec{q}_{\text{lo}}\preceq\vec{q}_{u}(t)\preceq\vec{q}_{\text{up}}$, where $\vec{q}_{\text{lo}}=(x_{\text{lo}}, y_{\text{lo}}, z_{\text{lo}})$, $\vec{q}_{\text{up}}=(x_{\text{up}}, y_{\text{up}}, z_{\text{up}})$ and $\preceq$ denotes the element-wise inequality. The initial and final locations of each DUE are indicated as $\vec{q}_{u}(0)=\vec{q}_{u}(I)$ and $\vec{q}_{u}(T_{u})=\vec{q}_{u}(D)$, respectively. Therefore, the trajectory\footnote{Note that this paper concentrates on RRM-aided performance optimization from the perspective of wireless communications, while DUEs are assumed to fly aloft with effective collision avoidance, powered by possible propulsion solutions, e.g., electric propulsion systems with motors. Thus, viewpoint of UAV's propulsion or hovering energy cost is beyond the scope of this paper.} of each DUE $u$ can be fully traced by $\vec{q}_{u}(t), \forall t\in[0, T_{u}]$.

\subsection{The RB Allocation Criterion}\label{SubSec_RBallocation}

To properly manage ICIs among $G$ GUEs, the following RB assignment criterion is adopted at all BSs \cite{mei2019cooperative, mei2019cellular}. The set $\mathcal{TI}_{b}(p)$ is defined to denote the first $p$-tier BSs that encompass a specific BS $b\in\mathcal{B}$ in the considered model, where $1\leq p\leq 3$ and $\mathcal{TI}_{b}(p)$ includes this focused BS. When arbitrary RB has been assigned to any GUE in the serving cells of BSs from $\mathcal{TI}_{b}(p)$, the focused BS $b$ should avoid allocating this RB to other GUEs in its corresponding cell.\footnote{In the case of sufficiently large $p$, the ICIs among all GUEs become ignorable, thanks to sufficient frequency reuse and severe terrestrial pathloss.} To ensure that the total RB resource is sufficient for all GUEs in cells of BSs from $\mathcal{TI}_{b}(p)$, the constraint $\sum_{\hat{b}\in\mathcal{TI}_{b}(p)}g_{\hat{b}}\leq K$ should hold, where $card(\mathcal{TI}_{b}(p))=3p^{2}+3p+1$ and $card(\cdot)$ indicates the cardinality of a set. In this regard, the focused BS $b$ cannot generate any ICIs to GUEs in the serving cells of BSs from $\mathcal{TI}_{b}(p)$. For GUEs outside the serving cells of BSs from $\mathcal{TI}_{b}(p)$, the potential ICIs caused by the focused BS $b$ are assumed to be negligible, due to severe terrestrial NLoS pathloss and shadowing. For each possible RB $k$, some BSs may already occupy it to serve GUEs in their corresponding cells. These BSs are recognized as the occupied BSs, which are denoted by the occupied BS set $\mathcal{B}_{o}^{k}\subset\mathcal{B}$. Furthermore, the set $\hat{\mathcal{B}}_{o}^{k}=\mathcal{B}\backslash\mathcal{B}_{o}^{k}$ includes all the potential BSs, where the RB $k$ is idle. For a specific RB $k$ assigned to serve a DUE, the corresponding associated BSs come from the potential set $\hat{\mathcal{B}}_{o}^{k}$, while all the non-associated co-channel interferences root from the occupied set $\mathcal{B}_{o}^{k}$. For a DUE $u$ associated with an RB $k$, it is supposed to be paired with all BSs in the potential set $\hat{\mathcal{B}}_{o}^{k}$, to take the advantage of macro-diversity gain. However, this may generate additional ICIs to GUEs in the serving cells of BSs from $\mathcal{TI}_{b\in\hat{\mathcal{B}}_{o}^{k}}(p)$. To avoid ICIs attenuating the receiving quality of existing GUEs over the same RB, a potential BS $b\in\hat{\mathcal{B}}_{o}^{k}$ can be allowed to pair DUE if and only if there are no other BSs applying RB $k$ in its first $p$-tier neighbours, i.e., 
\begin{equation}
	\mathcal{B}_{o}^{k}\cap\mathcal{TI}_{b\in\hat{\mathcal{B}}_{o}^{k}}(p)=\emptyset.
	\label{RB Allocation Criterion}
\end{equation}

Then, the available BS set $\breve{\mathcal{B}}_{o}^{k}\subset\hat{\mathcal{B}}_{o}^{k}$ is defined to denote the potential BSs satisfying (\ref{RB Allocation Criterion}). {For ease of understanding, Fig. \ref{RB_Diagram_Tier3} and Fig. \ref{RB_Allocation_Example} illustrate the first $p$-tier set for $p=1,2,3$, and one example of BS grouping for an RB $k$, respectively.}

\subsection{Channel Models}
\label{Sec_channnel_models}
\subsubsection{BS-to-GUE (B2G) Channel Model}
The B2G channel may include large-scale fading caused by NLoS pathloss and corresponding small-scale fading in practice. 
Specifically, the terrestrial small-scale fading component is denoted as $\vec{h}_{bg}\in\mathbb{C}^{1\times M}, \forall b\in\mathcal{B}, g\in\mathcal{G}$. Note that the modelling of $\vec{h}_{bg}$ is trivial for this paper, which means that $\vec{h}_{bg}$ can take form of any practical and feasible small-scale fading model, e.g., Rayleigh fading channel. In the section of numerical results, an example of terrestrial small-scale fading will be specified to perform the simulation.  
\subsubsection{BS-to-DUE ({B2D}) Channel Model}\label{LoS_Check}
According to 3GPP urban-macro (UMa) channel model \cite{ThreeGPP2017}, the expected {B2D} pathloss in dB can be expressed as $\text{PL}_{bu} = \text{Pr}_{\text{LoS}}\text{PL}_{\text{LoS}} + \text{Pr}_{\text{NLoS}}\text{PL}_{\text{NLoS}}$,
	\label{Probabilistic Path Loss Model}
where 
$\text{Pr}_{\text{LoS}}$ represents the occurrence probability of LoS link, $\text{Pr}_{\text{NLoS}}=1-\text{Pr}_{\text{LoS}}$ indicates that of NLoS channel,  and $\text{PL}_{\text{LoS}}$ and $\text{PL}_{\text{NLoS}}$ denote the pathlosses for LoS and NLoS links, respectively. Specifically, we have
\begin{equation}
	\text{Pr}_{\text{LoS}}=
	\begin{cases}
		\min\{\frac{\varepsilon_{1}}{r_{bu}},1\}
		\left[1-\exp\left(-\frac{r_{bu}}{\varepsilon_{2}}\right)\!\right]+\exp\left(-\frac{r_{bu}}{\varepsilon_{2}}\right),&22.5\text{m}<h\leq 100\text{m} \\
		1,&100\text{m}<h\leq 300\text{m}
	\end{cases},
	\label{Los_Nlos_Probability}
\end{equation}
\begin{equation}
	\text{PL}_{l}=
	\begin{cases}
		28.0+22\log_{10}\left(d_{bu}\right)+20\log_{10}\left(f_{c}\right),&l=\text{LoS} \\
		-17.5+\left[46-7\log_{10}\left(h\right)\right]
		\log_{10}\left(d_{bu}\right)+20\log_{10}\left(\frac{40\pi f_{c}}{3}\right),&l=\text{NLoS}
	\end{cases},
\end{equation}
in which $r_{bu}=\sqrt{d_{bu}^2-h^2}$, $\varepsilon_{1}=\max\{460\log_{10}\left(h\right)-700,18\}$, $\varepsilon_{2}=4300\log_{10}\left(h\right)-3800$, $f_{c}$ represents the carrier frequency and $d_{bu}=\vert\vert\vec{q}_{u}-\vec{q}_{b}\vert\vert_{2}$ calculates the Euclidean distance between DUE $u$ and ground BS $b$. Since the proposed design on beamforming vectors aims to be adaptive to arbitrary small-scale fading environment, {we denote $\vec{h}_{bu}\in\mathbb{C}^{1\times M}, \forall b\in\mathcal{B}, u\in\mathcal{U}$ as the small-scale fading component for {B2D} channels and pose no assumptions on its modelling, while an example of specific {B2D} small-scale fading model will be discussed in numerical result section.} 

\begin{figure}[htbp]
	\centering  
	\subfigcapskip=-.2cm
	\subfigure[2D distribution of local buildings and BSs]{
		\label{level.sub.1}
		\includegraphics[width=.35\linewidth]{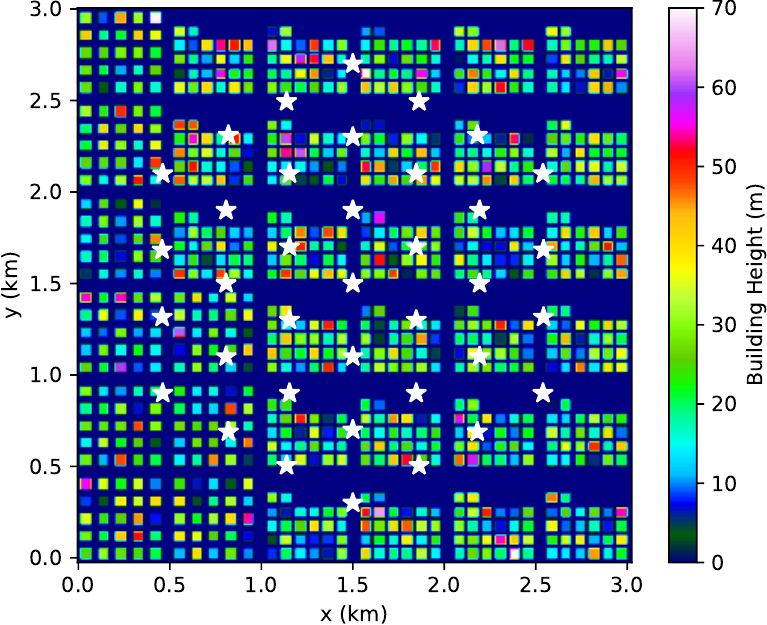}}
	\hspace{1.5cm}
	\subfigure[3D view of local building distribution]{
		\label{level.sub.2}
		\includegraphics[width=.33\linewidth]{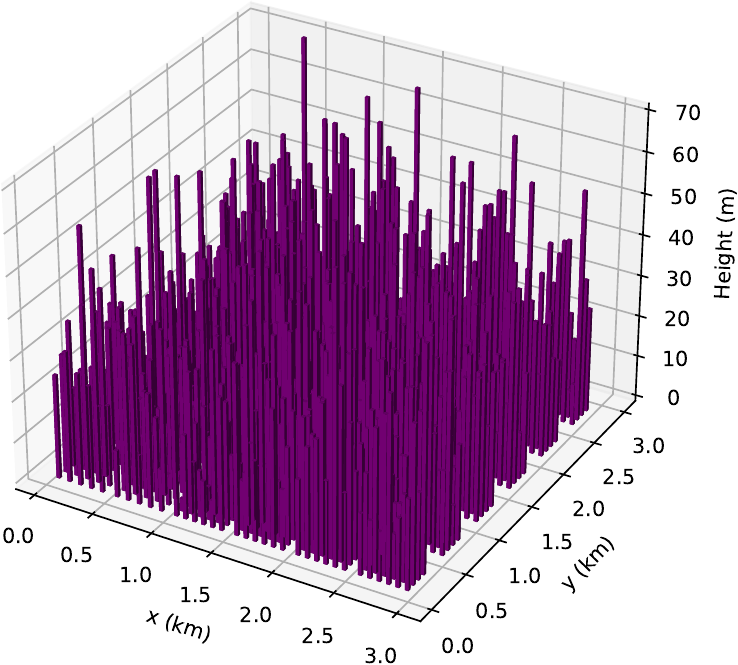}}
	\caption{The considered building distribution}
	\label{Building Distribution}
\end{figure}

To practically reflect the characteristics of {B2D} channels in the considered subregion, one realization of the statistical model suggested by the ITU \cite{series2013propagation}  is generated to formulate the building distribution, including high-rise structures' 2D locations on the ground and their corresponding heights. There are three key parameters in the ITU building distribution model \cite{li2022intelligent,li2022covertness}: 1) $\hat{\alpha}$ indicates the ratio of land region covered by buildings to the total land area; 2) $\hat{\beta}$ represents the mean of buildings per unit area; and 3) $\hat{\gamma}$ determines the distribution of building heights, which is typically following Rayleigh distribution with mean $\hat{\gamma}>0$. Note that the {B2D} pathlosses are modelled and tracked in terms of average large-scale channel gain via calculating the occurrence probabilities of LoS/NLoS links as depicted in (\ref{Los_Nlos_Probability}), in the vast majority of related literature. This kind of approach is more mathematically tractable, however, it can only reflect the ergodic characteristics of {B2D} channels over many realizations of building distribution {\cite{zeng2019path,zeng2021simultaneous,li2022path}}. On the contrary, in this paper, the occurrences of LoS/NLoS links are alternatively tracked via checking whether the line of {B2D} channel is blocked or not by any building, given one realization of ITU building distribution model.\footnote{Note that our approach is more practical because the building distribution of a subregion in real world can hardly vary over time, say, days even years. } Then, the corresponding type of large-scale pathloss can be accurately determined for each time of {B2D} channel regeneration. Fig. \ref{Building Distribution} illustrates the considered one realization of local building distribution in this paper, 
where 25 building clusters and 37 BSs are depicted in a square subregion with side length $D=3$ km, road width $\hat{D}=0.02$ km, $\hat{\alpha}=0.3$, $\hat{\beta}=103$ buildings/km$^2$ and $\hat{\gamma}=20$ m. With these parameter settings, the total amount of buildings is $\hat{\beta}D^{2}=927$ and the expected size of each building is $\hat{\alpha}/\hat{\beta}\approx0.003$ km$^2$. Besides, the maximum height of buildings is clipped to be under 70 m, and the locations of BSs are presented by white asterisks in Fig. \ref{level.sub.1}.%
\subsection{SINR at DUE}
Denote $C_{u}^{k}(t)\in\{0,1\}$ as the RB association indicator which means that DUE $u$ is occupying RB $k$ at time $t$ when $C_{u}^{k}(t)=1$, and $C_{u}^{k}(t)=0$ otherwise. Each DUE is assumed to occupy at most one single RB each time\footnote{In this paper, we focus on the scenario in which each DUE can only occupy one single RB each time. Integrating more sophisticated RB allocation approaches might be considered in our future works.}, then we have $\sum_{k=1}^{K}C_{u}^{k}(t)\leq1$.

If RB $k$ is feasible to be assigned to DUE $u$, i.e., $C_{u}^{k}(t)=1$, it has to satisfy the RB assignment criterion presented in Subsection \ref{SubSec_RBallocation}. Then, all BSs in the potential set $\hat{\mathcal{B}}_{o}^{k}$ meeting the regulation (\ref{RB Allocation Criterion}), i.e., $b\in\breve{\mathcal{B}}_{o}^{k}$, are recognized as the available BSs for DUE $u$, to take the advantage of macro-diversity gain. Besides, all BSs $b\in\mathcal{B}_{o}^{k}$ occupying the selected RB $k$ should be classified as the source of co-channel ICIs.
Thus, the received signal of DUE $u$ over RB $k$ at time $t$ can be given by
\begin{equation}
	y_{u}^{k}(t)=C_{u}^{k}(t)\left[\sum_{b\in\breve{\mathcal{B}}_{o}^{k}}\sqrt{10^{\frac{-\text{PL}_{l}}{10}}}\vec{h}_{bu}\vec{w}_{bu}x_{u}(t)\!\right.\left.+\!\sum_{b\in\mathcal{B}_{o}^{k}}\sqrt{10^{\frac{-\text{PL}_{l}}{10}}}\vec{h}_{bu}\vec{w}_{bg}x_{bg}(t)+n_{u}^{k}\right],
\end{equation}
where $\vec{w}_{bu}\in\mathbb{C}^{M\times 1}$ indicates the transmit beamforming vector at BS $b\in\breve{\mathcal{B}}_{o}^{k}$ for DUE $u$, $\vec{w}_{bg}\in\mathbb{C}^{M\times 1}$ represents the transmit beamforming vector at BS $b\in\mathcal{B}_{o}^{k}$ for corresponding GUEs, $x_{u}(t)\sim\mathcal{CN}(0,P)$ is the intended message\footnote{The available BSs are supposed to be able to cooperatively transmit the intended signal to DUE, managed by the central coordinator (to be introduced later) using, e.g., the cooperative beamforming technique \cite{mei2019cooperative}, while the procedure and overhead of cooperative transmissions are out the scope of this paper.} from BS $b$ to DUE $u$, $x_{bg}(t)\sim\mathcal{CN}(0,P)$ implies the signal for GUEs, and $n_{u}^{k}\sim\mathcal{CN}(0,\sigma^2)$ denotes the received additive complex Gaussian noise (AWGN) at DUE $u$. Note that the explicit type of large-scale fading between BS $b$ and DUE $u$ at time $t$, i.e., $l=\{\text{LoS},\text{NLoS}\}$, can be determined via checking possible blockages according to the considered one realization of local building distribution mentioned in Subsection \ref{LoS_Check}. Taking advantage of macro-diversity gain, all signals from the associated BS $b\in\breve{\mathcal{B}}_{o}^{k}$ are recognized as the legitimate in-phase information and thus can be added constructively at DUE $u$ \cite{mei2019cooperative,mudumbai2009distributed}. The channel state information (CSI) of $\vec{h}_{bu}, b\in\breve{\mathcal{B}}_{o}^{k}$ and $\vec{h}_{bg},b\in\mathcal{B}_{o}^{k}$ can be estimated via widely-applied MMSE-based methods. Unfortunately, the CSI {cannot} be perfectly obtained in practice, due to estimation error and/or feedback delay \cite{joudeh2016sum,choi2017joint}. Therefore, the imperfect CSI model on $\vec{h}_{bu}, b\in\breve{\mathcal{B}}_{o}^{k}$ is considered in this paper, expressed as
\begin{equation}
\vec{h}_{bu} = \sqrt{\rho}\vec{\ddot{h}}_{bu} + \sqrt{1 - \rho}\vec{\Delta},\label{imperfectCSI}
\end{equation}
where $\vec{\ddot{h}}_{bu}$ indicates the estimated CSI, $\vec{\Delta}\sim\mathcal{CN}(0,\boldsymbol{I})$ denotes the CSI estimation error vector and $\rho\in[0,1]$ is the correlation coefficient between $\vec{h}_{bu}$ and $\vec{\ddot{h}}_{bu}$. For an impractical case $\rho=1$, i.e., perfect CSI availability at the available BSs, maximum ratio transmission (MRT) precoding $\vec{w}_{bu}=\vec{h}_{bu}^{\dagger}/\Vert\vec{h}_{bu}\Vert$ is obviously the optimal option. However, $\vec{w}_{bu}$ should be designed as per the estimated CSI $\vec{\ddot{h}}_{bu}$, whose performance will be inevitably deteriorated due to the existence of CSI estimation error. 
Then, instantaneous SINR of DUE $u$ at time $t$ can be calculated as \cite{mei2019cooperative}
\begin{equation}
	\Gamma_{u}(t) =\sum_{k=1}^{K} \frac{ C_{u}^{k}(t)\left[\sum_{b\in\breve{\mathcal{B}}_{o}^{k}}\sqrt{P10^{\frac{-\text{PL}_{l}}{10}}}\vert\vec{h}_{bu}\vec{w}_{bu}\vert\right]^2}{I_{u}^{k}(t)+\sigma^2},\label{SINR_u}
\end{equation}
where $I_{u}^{k}(t)=\sum_{b\in\mathcal{B}_{o}^{k}}P10^{\frac{-\text{PL}_{l}}{10}}\vert\vec{h}_{bu}\vec{w}_{bg}\vert^2$ means ICIs introduced by co-channel BSs from $\mathcal{B}_{o}^{k}$. 

\subsection{Problem Formulation}
It is clear that the received SINR of DUE $u$ at time $t$, i.e., formula (\ref{SINR_u}), is a random variable {because of the randomness introduced by, e.g., small-scale fadings and RB allocation.} Specifically, the RB allocation affects $\Gamma_{u}(t)$ in terms of how many available BSs and interfering BSs will be involved, i.e., $card(\breve{\mathcal{B}}_{o}^{k})$ and $card(\mathcal{B}_{o}^{k})$, respectively. Then, with given RB allocation, the transmit beamforming vector $\vec{w}_{bu}$ should be designed to adapt to $\vec{h}_{bu}$.
Hence, the corresponding transmission outage probability (TOP) is formulated as a function of $C_{u}^{k}(t)$ and $\vec{w}_{bu}$, given by
\begin{equation}
TOP_{u}\{C_{u}^{k}(t), \vec{w}_{bu}\} = \Pr\left[\Gamma_{u}(t)<\Gamma_{th}\right],
\end{equation}
where $\Pr$ outputs the probability calculated with respect to (w.r.t.) the aforementioned small-scale fadings and {B2D} transmit beamforming vector, for given RB allocation. Then, the EOD of DUE $u$ travelling with trajectory $\vec{q}_{u}(t), \forall t\in[0, T_{u}]$ from $\vec{q}_{u}(I)$ to $\vec{q}_{u}(D)$ can be expressed as
\begin{equation}
EOD_{u}\{C_{u}^{k}(t), \vec{w}_{bu}\}
=\int^{T_{u}}_{0}TOP_{u}\{C_{u}^{k}(t), \vec{w}_{bu}\}dt.\label{EOD_{u}}
\end{equation}

This paper assumes that DUEs move with known trajectories $\vec{q}_{u}(t), \forall u\in\mathcal{U}, t\in[0, T_{u}]$ and constant velocity $V_{u}$, then $T_{u}$ in (\ref{EOD_{u}}) can be implied as a fixed parameter posing no impacts on the overall integral.\footnote{Note that RB allocation and beamforming design are independent of trajectories, which means that the proposed solution is suitable for arbitrary UAV trajectory. This setup can be justified by the following facts: 1) as elaborated in Subsection \ref{SubSec_RBallocation}, RB coordination for DUEs depends on the current RB possession of each BS and has nothing to do with DUEs' mobility; and 2) beamforming design depends on the estimated CSI which is related to the corresponding modelling of small-scale fading. Therefore, trajectory planning task is trivial in the considered system model and thus excluded from this paper.} Hence, the EOD of arbitrary DUE $u$ is fully determined by $C_{u}^{k}(t)$ and $\vec{w}_{bu}$. Without loss of generality, in the following contents of this paper, a specific DUE in Fig. 1 is focused to evaluate our proposed scheme which can be easily applied to other DUEs with orthogonal RB assignment. For enhancing the down-link transmission quality of DUE across its travelling trajectory $\vec{q}_{u}(t)$, this paper focuses on minimizing its EOD. Then, the corresponding optimization problem can be stated as 

\begin{subequations}
	\setstretch{.2}
	\begin{align}
	(\text{P}1):&\min\limits_{C_{u}^{k}(t), \vec{w}_{bu}}EOD_{u}\{C_{u}^{k}(t), \vec{w}_{bu}\},\label{ProposedProblem}\\
	\text{s.t.}&\sum_{k=1}^{K}C_{u}^{k}(t)\leq1, \forall t\in[0,T_{u}], \label{Constraint1} \\ 
	&\vert\vert\vec{w}_{bu}\vert\vert^2=1, \forall b\in\breve{\mathcal{B}}_{o}^{k}, \forall t\in[0,T_{u}],\label{Constraint2} \\
	&C_{u}^{k}(t)\in\{0,1\},\forall k\in\mathcal{K}, \forall t\in[0,T_{u}],\label{Constraint3}\\
	&\vec{q}_{\text{lo}}\preceq\vec{q}_{u}(t)\preceq\vec{q}_{\text{up}}, \forall t\in[0,T_{u}].\label{Constraint4}
	\end{align}
\end{subequations}

The constraint (\ref{Constraint1}) makes sure that the DUE can at most occupy one single RB each time. The constraint (\ref{Constraint2}) is the normalization requirement for transmit beamforming vector, which ensures that the transmit power of each available BS $b\in\breve{\mathcal{B}}_{o}^{k}$ equals $P$. The constraint (\ref{Constraint3}) indicates that $C_{u}^{k}(t)$ is a binary variable. {The constraint (\ref{Constraint4}) claims that DUE's trajectory should remain in the considered subregion.} 

It is extremely challenging to solve the proposed optimization problem (P1), given the listed constraints. The main difficulties can be portrayed as follows: 1) the closed-form expression of $EOD_{u}\{C_{u}^{k}(t), \vec{w}_{bu}\}$ should be derived, which is extraordinarily sophisticated, if not impossible; 2) the variations of LoS/NLoS pathloss, small-scale fading $\vec{h}_{bu}$ and the B2G transmit beamforming vector $\vec{w}_{bg}$ should be taken into consideration, which are dynamic over time horizon and dependent on their modellings; 3) even given the closed-form expression of the optimization object (\ref{ProposedProblem}) and the perfect knowledge of the considered cellular-connected UAV network, it is still mathematically inefficient to be tackled for the non-convexity of mix-integer constraint (\ref{Constraint3}) and that of the optimization object (\ref{ProposedProblem}) w.r.t. $C_{u}^{k}(t)$ and $\vec{w}_{bu}$. {Fortunately, DRL is famous for being able to learn patterns from unknown environments in a trial-and-error fashion and thus can help solve sophisticated optimization problems via inherently maximizing its long-term return of raw optimization objective. Thus, this paper resorts to initiating a DRL solution to tackle (P1).} 
\section{The Proposed DRL-Aided Algorithm}
\label{d3qn_ddpg_section_the_proposed_algorithm}
\subsection{The Formulation of MDP}
To design DRL-based solution for the proposed optimization problem (P1), the first step is to formulate (P1) into MDP which is based on discrete time slots \cite{xiao2021uav}. The length of time slot is defined as $\delta_{u}$ for the considered model and thus the number of time slots equals $N_{u}=T_{u}/\delta_{u}$ for the DUE $u$. Note that the duration of time slot $\delta_{u}$ should be picked as small as possible, to achieve that the distances between the DUE and BSs remain approximately constant in each time slot. 
In this regard, the EOD expression can be rewritten as
\begin{equation}
EOD_{u}\{C_{u}^{k}(n), \vec{w}_{bu}\}
\approx \sum_{n=1}^{N_{u}}\delta_{u}TOP_{u}\{C_{u}^{k}(n), \vec{w}_{bu}\}.\label{EOD_{u}_Discrete}
\end{equation}

However, even with given $C_{u}^{k}(n)$, the closed-form expression of the transmission outage probability $TOP_{u}\{C_{u}^{k}(n), \vec{w}_{bu}\}$ is still difficult to be derived, for its complex formulation and the lack of designed {B2D} transmit beamforming vector $\vec{w}_{bu}$. Alternatively, this challenge can be circumvented via numerical evaluation on the raw measurements of received signals at the DUE. The reason is that, compared to the length of time slot $\delta_{u}$ (typically on the magnitude of seconds), the length of channel coherence blocks (typically on the magnitude within milliseconds) is relatively insignificant \cite{zeng2021simultaneous, li2022path}. Then, provided with $C_{u}^{k}(n)$ for a time slot $n$, the indicator of TOP can be defined as $ITOP_{u}\{C_{u}^{k}(n), \vec{w}_{bu}(n,i);\hat{h}(n,i)\}=1$ in the case of $\Gamma_{u}(n,i)<\Gamma_{th}$, and $ITOP_{u}\{C_{u}^{k}(n), \vec{w}_{bu}(n,i);\hat{h}(n,i)\}=0$ otherwise, where $\hat{h}(n,i)$ and $\vec{w}_{bu}(n,i)$ indicate one realization of small-scale fadings and that of corresponding beamforming vector, respectively.

Then, the corresponding TOP can be calculated as
\begin{equation}
	TOP_{u}\{C_{u}^{k}(n), \vec{w}_{bu}\}=\mathbb{E}_{\hat{h},\vec{w}}\left[ITOP_{u}\{C_{u}^{k}(n), \vec{w}_{bu}(n,i);\hat{h}(n,i)\}\right].\label{TOP_Statistical}
\end{equation}

To realize the average calculation $\mathbb{E}_{\hat{h},\vec{w}}$ over $\hat{h}$  and $\vec{w}$ in (\ref{TOP_Statistical}), $\varsigma$ times of SINR measurement should be performed.\footnote{The existing soft handover technique, accompanied by reference signal received power (RSRP) and reference signal received quality (RSRQ) reports, can be applied to help complete this kind of task \cite{zeng2021simultaneous}.}
Furthermore, the arithmetic TOP of the DUE $u$ can be expressed as
\begin{equation}
	\bar{TOP}_{u}\{C_{u}^{k}(n), \vec{w}_{bu}\}=\frac{1}{\varsigma}\sum_{i=1}^{\varsigma}ITOP_{u}\{C_{u}^{k}(n), \vec{w}_{bu}(n,i);\hat{h}(n, i)\}.\label{TOP_arithmetic}
\end{equation}
When sufficiently large amount of SINR measurements is performed, i.e., $\varsigma\gg1$, the statistical average in (\ref{TOP_Statistical}) can be alternatively replaced by its arithmetic counterpart in (\ref{TOP_arithmetic}).\footnote{In the case of $\varsigma\rightarrow+\infty$, $\lim_{\varsigma\rightarrow+\infty}\bar{TOP}_{u}\{C_{u}^{k}(n), \vec{w}_{bu}\}=TOP_{u}\{ C_{u}^{k}(n), \vec{w}_{bu}\}$ can be guaranteed theoretically.} Thereafter, the EOD expression in (\ref{EOD_{u}_Discrete}) can be modified as
\begin{equation}
EOD_{u}\{C_{u}^{k}(n), \vec{w}_{bu}\}\approx\! \sum_{n=1}^{N_{u}}\sum_{i=1}^{\varsigma}\frac{\delta_{u}}{\varsigma}ITOP_{u}\{ C_{u}^{k}(n), \vec{w}_{bu}(n,i);\hat{h}(n, i)\}.\label{EOD_{u}_Final}
\end{equation}

Then, the original optimization problem (P1) can be approximately revised as

\begin{subequations}\label{formulated_RRM_problem}
	\setstretch{.2}
	\begin{align}
	(\text{P}2):&\min\limits_{C_{u}^{k}(n), \vec{w}_{bu}(n,i)}\sum_{n=1}^{N_{u}}\sum_{i=1}^{\varsigma}\frac{\delta_{u}}{\varsigma}
	ITOP_{u}\{C_{u}^{k}(n), \vec{w}_{bu}(n,i);\hat{h}(n, i)\},\label{ProposedProblem2}\\
	\text{s.t.}&\sum_{k=1}^{K}C_{u}^{k}(n)\leq1, \forall n\in[1,N_{u}], \label{Constraint21}\\ 
	&\vert\vert\vec{w}_{bu}(n,i)\vert\vert^2=1, \forall b\in\breve{\mathcal{B}}_{o}^{k}, \forall n\in[1,N_{u}],\label{Constraint22} \\
	&C_{u}^{k}(n)\in\{0,1\},\forall k\in\mathcal{K}, \forall n\in[1,N_{u}],\label{Constraint23}\\
	&\vec{q}_{\text{lo}}\preceq\vec{q}_{u}(n)\preceq\vec{q}_{\text{up}}, \forall t\in[1,N_{u}].\label{Constraint24}
	\end{align}
\end{subequations}

{Inspired by cloud radio access network (C-RAN) \cite{kadan2021theoretical} and cell-free (CF) distributed MIMO \cite{elhoushy2021cell}, the terrestrial BSs are controlled by a central coordinator (C2)\footnote{The C2 is typically hosted in the edge cloud platform, and thereby provides high-performance computing and centralized signal processing for a large number of UEs' data.} via high-speed fronthaul links, e.g., optical fiber, to realize the joint RB allocation and beamforming design task.} Once the DUE $u$ registers into the cellular network, the C2 will first check the overall RB availability of all BSs, after which a map of RB possession (RBP) formulated as a 2D matrix $\mathcal{C}(n)=[C_{b}^{k}(n)]_{b\times k}$ will be generated. Note that $C_{b}^{k}(n)=1$ if RB $k$ is occupied by BS $b$ at time slot $n$ and $C_{b}^{k}(n)=0$ otherwise. Then, for each RB $k$, following the RB allocation criterion presented in Subsection \ref{SubSec_RBallocation}, the corresponding occupied set $\mathcal{B}_{o}^{k}$, the potential set $\hat{\mathcal{B}}_{o}^{k}$ and the available set $\breve{\mathcal{B}}_{o}^{k}$ can be determined. Taking the advantage of macro-diversity gain, the C2 will assign all available BSs $b\in\breve{\mathcal{B}}_{o}^{k}$ to serve the DUE cooperatively. Note that $\mathcal{C}(n)$ remains constant within each time slot and varies among different time slots\footnote{To avoid frequent handover, the selected RB $k$ is considered as unchanged within each time slot.}, capturing the dynamics of RBP at terrestrial BSs. For each time slot, the current location of DUE $\vec{q}_{u}(n)$ is observable. Then, the large-scale fading distribution between the DUE and BSs can be traced, via checking the potential blockages between the DUE and each BS according to the local building distribution as mentioned in Subsection \ref{Sec_channnel_models}. From the point view on SINR in (\ref{SINR_u}), the allocated RB $k$ serving the DUE can affect the value of SINR in terms of how many desired channels and interfering links are introduced. Hence, the selection of RB resource can inherently impact the EOD performance and should be delicately assigned. Next, given specific RB within each time slot, the beamforming strategy adapting to the time-varying small-scale fading component can further affect the EOD performance. 

To handle the aforementioned two-step process, a hybrid D3QN-TD3 algorithm\footnote{Please note that the proposed DRL-aided solution is trained online at the C2, rather than each BS.} is proposed, in which an outer MDP is formulated for the D3QN agent while an inner MDP is forged for the TD3 agent. Specifically, the D3QN determines which RB should be selected for each time slot and the TD3 outputs the proper beamforming vector for each link between the DUE and BSs in the available BS set. Furthermore, the considered cellular-connected UAV network is divided into the outer environment and the inner environment. For time slot $n$, the DUE's location $\vec{q}_{u}(n)$ and the RBP map $\mathcal{C}(n)$ can be observed from the outer environment. The inner environment is defined to reflect the time-varying characteristic of small-scale fading, which is dependent on the outer environment. The reason roots from that the {B2D} channel's small-scale fading component is subject to the corresponding experienced type of pathloss in practice, i.e., LoS or NLoS. 

\subsection{Description of the Hybrid D3QN-TD3 Solution}
To derive a flexible solution solving (P2) in a dynamic RBP and time-varying small-scale fading scenario, both D3QN and TD3 networks in the proposed hybrid D3QN-TD3 algorithm are trained interactively. Specifically, the D3QN network maps the outer state and the RB selection into Q values\footnote{The Q value $Q_{\pi}(s,a)$, i.e., state-action value, derives the discounted accumulated-rewards and reflects the long-term return after executing action $a$ over state $s$ following current action selection policy $\pi$, which is the typical metric to be maximized within DRL regime \cite{wang2016dueling, lillicrap2015continuous}.}, while the actor of TD3 agent transforms the inner state to beamforming vector and the critic of TD3 network evaluates the corresponding Q values. 

\subsubsection{D3QN}

\begin{figure}[htbp]
	\centering
	\includegraphics[width=1\linewidth]{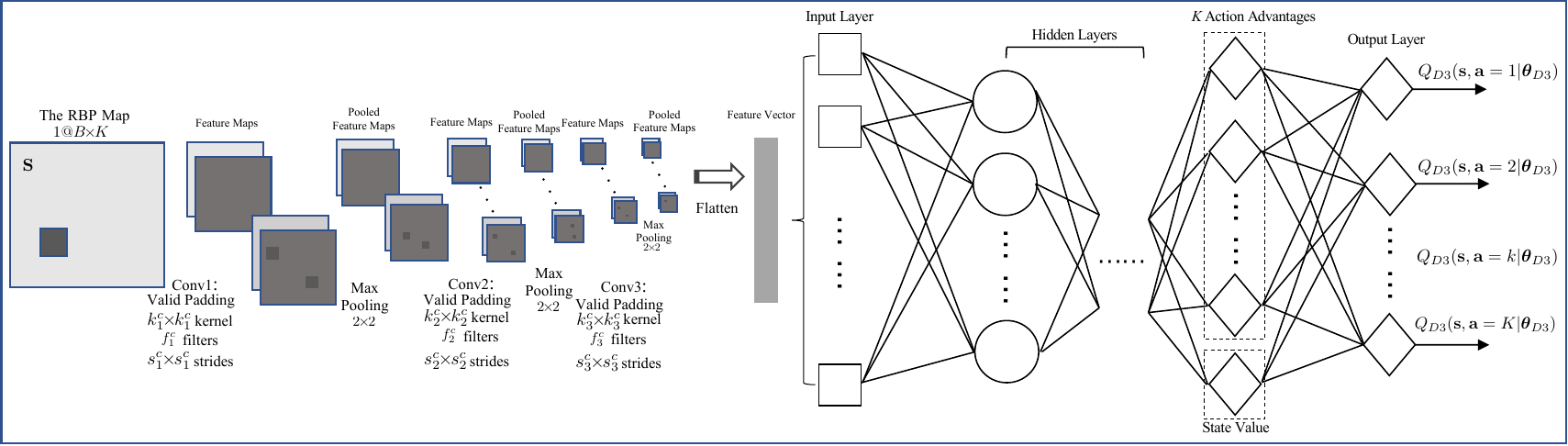}
	\caption{Architecture of CNN-attached duelling DQN}
	\label{Duelling DQN Diagram}
\end{figure}

To tackle the RB allocation problem, state-of-the-art deep Q network (DQN) with duelling architecture \cite{wang2016dueling} will be invoked to approximate Q function for the outer MDP. {In contrast to the original DQN method, the duelling DQN explicitly separates the estimations of state value and the corresponding action advantages into two independent streams. For clarity, Fig. \ref{Duelling DQN Diagram} depicts the network architecture of duelling DQN, where Q values are calculated via taking aggregation of the estimated state value and state-dependent action advantages. The duelling architecture, i.e., the two steams of neural networks, can help approximate Q values more robustly and efficiently when Q values of various actions with the same state are indistinguishable.}

The outer MDP for the D3QN agent can be formulated as follows. The outer state $\mathbf{s}$ is the observed RBP map $\mathcal{C}(n)$\footnote{The transition of RBP map is stochastic and can be observed from the outer environment, which means that the D3QN learning process is model-free.}, while the outer action $\mathbf{a}$ refers to the selected RB $k^{*}=\underset{k}{\arg}\{C_{u}^{k}(n)=1\}$. When the dimensionality of $\mathcal{C}(n)$ is significant, the computation and training burdens could be unbearable if the RBP map is {simply} flattened and then fed to the input layer of D3QN. To circumvent this issue, a convolutional neural network (CNN) is attached to the D3QN, for efficiently capturing the features of the RBP map and compressing the data fed into the D3QN. Specifically, the CNN contains three convolutional layers, i.e., Conv1, Conv2 and Conv3, as depicted in Fig. \ref{Duelling DQN Diagram} where the corresponding size of kernel, amount of filter and size of stride are denoted. Following each convolutional layer, a standard max pooling layer with pool size $2\times2$ and stride $2\times2$ is invoked. 
In the end, the pooled feature maps will be flattened into a vector which will then be fed into the input layer of D3QN. The considered optimization problem is fully determined by the value of SINR, given SINR threshold. In other word, larger available BS set and smaller occupied BS set are favourable to minimize the EOD. For outer state $\mathbf{s}$ and the selected outer action $\mathbf{a}$, the corresponding available BS set $\breve{\mathcal{B}}_{o}^{k^{*}}$ and the occupied BS set $\mathcal{B}_{o}^{k^{*}}$ can be determined according to Subsection \ref{SubSec_RBallocation}. Then, the outer reward function is defined as
\begin{equation}
\mathbf{r} = \frac{card(\breve{\mathcal{B}}_{o}^{k^{*}})}{card(\breve{\mathcal{B}}_{o}^{k^{*}})+card(\mathcal{B}_{o}^{k^{*}})}.\label{reward_outer}
\end{equation}
The designed outer reward function (\ref{reward_outer}) infers that the selected RB $k^{*}$ resulting in larger available BS set and smaller occupied BS set is more favourable.
Given the formulation of outer MDP, the duelling DQN is invoked to approximate $Q_{D3}(\mathbf{s},\mathbf{a}\vert\boldsymbol{\theta}_{D3})$ where $\boldsymbol{\theta}_{D3}$ represents the parameter vector of D3QN network. The D3QN network is trained to minimize its loss function via the gradient descent updating rule, shown as
\begin{equation}
		\boldsymbol{\theta}_{D3}(t+1)=\boldsymbol{\theta}_{D3}(t) - \alpha_{D3}\nabla_{\boldsymbol{\theta}_{D3}}loss(\boldsymbol{\theta}_{D3}),\label{descent_d3qn}
\end{equation}
where $\alpha_{D3}$ denotes the learning rate and $\nabla_{\boldsymbol{\theta}_{D3}}loss(\boldsymbol{\theta}_{D3})$ represents the gradient of the D3QN network's loss function w.r.t. $\boldsymbol{\theta}_{D3}$. For a mini-batch 
of $N_{D3}$ transitions randomly sampled from the outer replay buffer, the mean-square loss function in (\ref{descent_d3qn}) is defined as
\begin{equation}
loss(\boldsymbol{\theta}_{D3})=\frac{1}{N_{D3}}\sum_{t=1}^{N_{D3}}\left[y_{t}-Q_{D3}(\mathbf{s}_{t},\mathbf{a}_{t}\vert\boldsymbol{\theta}_{D3})\right]^2, \label{loss_function_d3qn}
\end{equation} 
where $y_{t} = \mathbf{r}_{t} +\gamma Q_{D3}(\mathbf{s}_{t+1},\mathbf{a}_{t+1}^{*}\vert\boldsymbol{\theta}_{D3}^{-})$ and $\boldsymbol{\theta}_{D3}^{-}$ indicates the parameter vector of target D3QN network. Note that the optimal outer action for the next outer state $\mathbf{s}_{t+1}$ is selected by the D3QN network instead of the target D3QN network, given by
\begin{equation}
\mathbf{a}_{t+1}^{*}=\underset{\mathbf{a}_{t+1}}{\arg\max} Q_{D3}(\mathbf{s}_{t+1},\mathbf{a}_{t+1}\vert\boldsymbol{\theta}_{D3}).
\end{equation} 
In this manner, the bootstrapping outer action is evaluated by the target D3QN network while the selection of outer action is achieved by the D3QN network, completing the double Q learning procedure.
Applying double Q learning method to separate action selection and bootstrapping evaluation into two networks can help address the overestimation bias issue introduced by the $\max$ operator in calculating the loss function. After several steps of updating the D3QN network, the target D3QN network will be synchronized to the D3QN network via letting $\boldsymbol{\theta}^{-}_{D3}=\boldsymbol{\theta}_{D3}$. 

Given outer state $\mathbf{s}$, the outer action selection strategy applied by the D3QN agent follows the popular $\epsilon$-greedy policy, shown as
\begin{equation}
	\mathbf{a}=
	\begin{cases}
		\text{randi}(K), & \text{with probability} \hspace{.1cm} \epsilon \\
		\underset{k=1,\dots,K}{\arg\max}Q_{D3}(\mathbf{s},k\vert\boldsymbol{\theta}_{D3}), & \text{otherwise}
	\end{cases},
	\label{epsilon_greedy_action_selection_policy}
\end{equation}
where {$\text{randi}(K)$ outputs a random integer from the range $\left[1,K\right]$ and} the exploration factor $\epsilon\in[0,1]$ is used to balance exploration and exploitation during training. Specifically, {greater $\epsilon$ encourages D3QN agent to explore the outer action space, while smaller $\epsilon$ results in more frequent exploitation of learned knowledge. Commonly, $\epsilon$ keeps annealing alongside the learning process,} steering D3QN agent from more frequent exploration to a higher probability of exploitation.

\subsubsection{TD3}
For each time slot $n$, the D3QN agent observes the outer environment, from which it obtains  the DUE's location $\vec{q}_{u}(n)$ and the RBP map $\mathcal{C}(n)$. Then, the D3QN agent selects the outer action, i.e., the RB $k^{*}$. With the selected RB and the current RBP map, the corresponding set of available BSs $\breve{\mathcal{B}}_{o}^{k^{*}}$ can be determined. To reduce the overheads of CSI estimation and inner reward feedback, a random BS out of the current available BSs will be selected by the C2 to perform beamforming optimization. Thereafter, the type of large-scale fading between the DUE and the chosen available BS can be obtained. 
Then, the inner MDP for the TD3 network can be formulated as follows. Each inner state $\hat{\mathbf{s}}$ consists of a list of estimated CSI $\vec{\ddot{h}}_{bu}(n,i)$ and its corresponding type of LoS or NLoS. 
It is well known that artificial neural networks (ANNs) only accept real numbers as their inputs, rather than complex values. To circumvent this problem, the complex-value estimated CSI $\vec{\ddot{h}}_{bu}(n,i)$ will be transferred into a flatten layer which decouples the complex value and reshapes its real and imagery parts into a real-value vector. However, the inner state $\hat{\mathbf{s}}$ is dominated by the flattened CSI, while only one dimension is left for the indicator of pathloss type, which raises the issue of dimension imbalance. To solve this, the dimension for pathloss type indicator will be expanded from 1 to $M$ via duplicating the pathloss type indicator into $M$ copies, making it comparable to the dimension of flattened CSI. Each possible inner action $\hat{\mathbf{a}}$ generated from the actor network is a vector of real-value numbers, which will be reshaped into a normalized complex-value vector to construct the corresponding beamforming vector $\vec{w}_{bu}(n,i)$. The transitions of inner states are determined by the experienced small-scale fading model. The inner reward function evaluates how good the selected inner action is for each time of state transition. To reflect the quality of selected inner action, the inner reward function is defined as 
\begin{equation}
	\hat{\mathbf{r}} = \frac{\vert\vec{h}_{bu}(n,i)\vec{w}_{bu}(n,i)\vert^2}{\Vert\vec{\ddot{h}}_{bu}(n,i)\Vert^2}.
	\label{inner_reward_function}
\end{equation}

\begin{figure}[htbp]
	\centering  
	\subfigcapskip=-.1cm
	\subfigure[Layout of the actor network]{
		\label{Actor}
		\includegraphics[width=.48\linewidth]{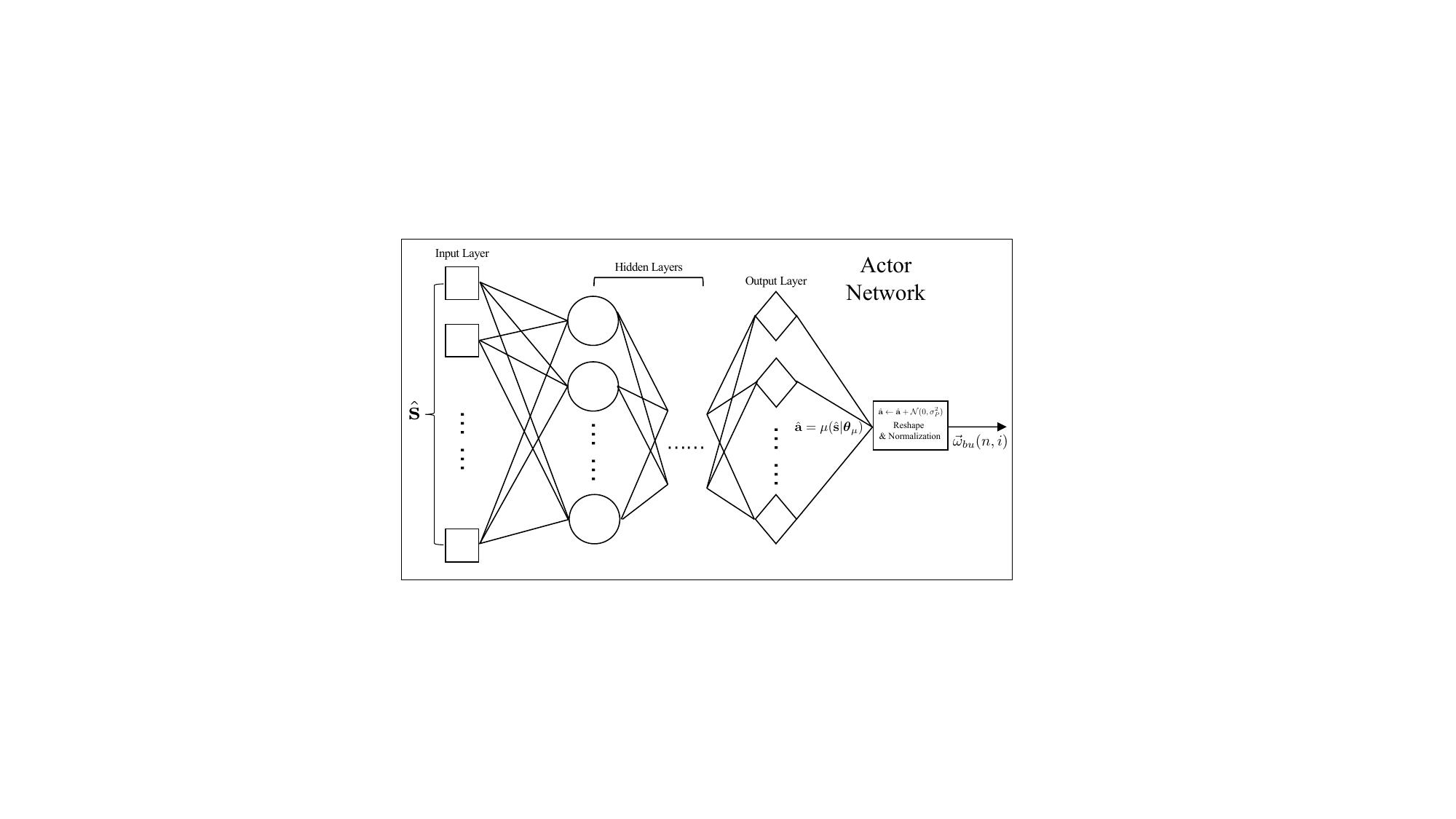}}
	\subfigure[Layout of the critic network]{
		\label{Critic}
		\includegraphics[width=.48\linewidth]{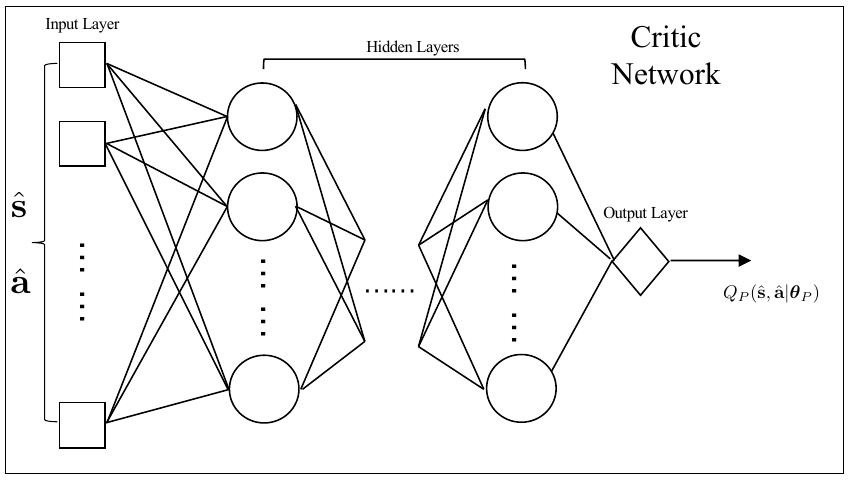}}
	\caption{Architecture of TD3 network}
	\label{TD3 Network}
\end{figure}

TD3 method belongs to actor-critic algorithms \cite{li2022covertness}, in which the critic network learns Q function approximation $Q_{P}(\hat{\mathbf{s}},\hat{\mathbf{a}}\vert\boldsymbol{\theta}_{P})$ and the actor network is the policy generator approximating the action $\mu(\hat{\mathbf{s}}\vert\boldsymbol{\theta}_{\mu})$. {Herein,} $\boldsymbol{\theta}_{P}$ and $\boldsymbol{\theta}_{\mu}$ denote the parameter vectors of critic and actor networks, respectively. {As illustrated in Fig. \ref{TD3 Network}, both actor and critic networks are constructed in line with popular actor-critic implementation in current literature \cite{lillicrap2015continuous,mnih2016asynchronous,schulman2017proximal}. In specific,} the actor network takes the inner state as its input and generates deterministic continuous action as its output, unlike DQN-related methods that output a probability distribution over discrete action space. Furthermore, the inner action generated by the actor network will be leveraged to the input layer of the critic network together with the current inner state. Then, the corresponding state-action value will be generated at the output layer of the critic network. {Please note that the actor network is invoked to approximate the inner action, which helps avoid an exhaustive search of the optimal inner action maximizing the Q function given the next inner state.}

The gradient descent updating on the twin critic networks can be given by
\begin{equation}
	\boldsymbol{\theta}_{P_{\hat{j}}}(t+1)=\boldsymbol{\theta}_{P_{\hat{j}}}(t) - \alpha_{Pc}\nabla_{\boldsymbol{\theta}_{P_{\hat{j}}}}loss(\boldsymbol{\theta}_{P_{\hat{j}}}),\label{descent_td3_critic}
\end{equation}
where $\alpha_{Pc}$ indicates the learning rate, $\nabla_{\boldsymbol{\theta}_{P_{\hat{j}}}}loss(\boldsymbol{\theta}_{P_{\hat{j}}})$ denotes the gradient of critic network's loss function w.r.t. $\boldsymbol{\theta}_{P_{\hat{j}}}$ and $\hat{j}\in\{1, 2\}$ is defined to distinguish the twin critics. Besides, the corresponding mean-square loss function is defined as
\begin{equation}
	loss(\boldsymbol{\theta}_{P_{\hat{j}}})=\frac{1}{N_{P}}\sum_{t=1}^{N_{P}}\left[\hat{y}_{t}-Q_{P}(\hat{\mathbf{s}}_{t},\hat{\mathbf{a}}_{t}\vert\boldsymbol{\theta}_{P_{\hat{j}}})\right]^2, \label{loss_function_td3}
\end{equation} 
where 
\begin{equation}
	\hat{y}_{t}=\hat{\mathbf{r}}_{t} +\gamma\underset{\hat{j}=1,2}{\min} Q_{P}[\hat{\mathbf{s}}_{t+1},\mu(\hat{\mathbf{s}}_{t + 1}\vert\boldsymbol{\theta}_{\mu}^{-}) + \mathcal{N}^{-}\vert\boldsymbol{\theta}_{P_{\hat{j}}}^{-}]
	\label{loss_function_clipped_double_Q_Chapter_hybrid_d3qn_td3}
\end{equation} 
represents the target Q value, $N_{P}$ is from a mini-batch of $N_{P}$ transitions extracted from the inner replay buffer, and $\boldsymbol{\theta}_{P_{\hat{j}}}^{-}$, $\boldsymbol{\theta}_{\mu}^{-}$ and $\mathcal{N}^{-}$ denote the parameters of target critic network, those of target actor network and additive noise for target actor network, respectively. Note that the operator $\min$ in (\ref{loss_function_clipped_double_Q_Chapter_hybrid_d3qn_td3}) and $\mathcal{N}^{-}$ are posed for accomplishing clipped double Q learning and target policy smoothing, respectively.

Moreover, the actor network aims to maximize its expected return, defined as
\begin{equation}
	J(\boldsymbol{\theta}) = \mathbb{E}_{\hat{\mathbf{s}}_{t}}\{Q[\hat{\mathbf{s}}_{t},\mu(\hat{\mathbf{s}}_{t}\vert\boldsymbol{\theta}_{\mu})\vert\boldsymbol{\theta}_{P}]\},
\end{equation}
of which the derivative w.r.t. $\boldsymbol{\theta}_{\mu}$ can be calculated with help of the chain rule, shown as
\begin{align}
	\nabla_{\boldsymbol{\theta}_{\mu}}J(\boldsymbol{\theta}) &\approx \mathbb{E}_{\hat{\mathbf{s}}_{t}}\{\nabla_{\boldsymbol{\theta}_{\mu}}Q[\hat{\mathbf{s}}_{t},\mu(\hat{\mathbf{s}}_{t}\vert\boldsymbol{\theta}_{\mu})\vert\boldsymbol{\theta}_{P}]\}\nonumber\\
	&=\frac{1}{N_{P}}\sum_{t=1}^{N_{P}}\nabla_{a}Q_{P}(\hat{\mathbf{s}}_{t}, a\vert\boldsymbol{\theta}_{P_{1}})\nabla_{\boldsymbol{\theta}_{\mu}}\mu(\hat{\mathbf{s}}_{t}\vert\boldsymbol{\theta}_{\mu}),
\end{align}
in which the critic 1 is anchored by the chain rule for simplicity.

Then, the gradient ascent updating of the actor network can be expressed as
\begin{equation}
	\boldsymbol{\theta}_{\mu}(t+1)=\boldsymbol{\theta}_{\mu}(t) + \alpha_{Pa}\nabla_{\boldsymbol{\theta}_{\mu}}J(\boldsymbol{\theta}),\label{ascent_td3_actor}
\end{equation}
where $\alpha_{Pa}$ is the learning rate of the actor network. Moreover, to complete the delayed policy update procedure, the actor, target actor and the twin target critics will be updated less frequently than the twin critics, via updating the target networks every $N_{pud}$ times the twin critics are trained. 

Furthermore, the Polyak averaging updates for the target critic and actor networks are applied to enhance the stability of learning, given by
\begin{equation}
	\boldsymbol{\theta}_{P_{\hat{j}}}^{-} \leftarrow \tau\boldsymbol{\theta}_{P_{\hat{j}}} + (1 - \tau)\boldsymbol{\theta}_{P_{\hat{j}}}^{-},\label{Polyak_critic}
\end{equation}
\begin{equation}
	\boldsymbol{\theta}_{\mu}^{-} \leftarrow \tau\boldsymbol{\theta}_{\mu} + (1 - \tau)\boldsymbol{\theta}_{\mu}^{-},\label{Polyak_actor}
\end{equation}
respectively, where $\tau$ is the interpolation factor in Polyak averaging method for updating target networks and it is usually set to be close to zero, i.e., $\tau\ll1$.

Different from probabilistic action selection policy on discrete actions for D3QN agent, exploration on continuous actions for TD3 agent can be realized via adding noise sampled from a noise process $\mathcal{N}$ to the actor network, i.e., $\hat{\mathbf{a}} \leftarrow \hat{\mathbf{a}} + \mathcal{N}$,
where $\mathcal{N}$ can be chosen to adapt to the inner environment \cite{lillicrap2015continuous}. For simplicity, zero-mean Normal noise with variance $\sigma_{P}^{2}$ is applied to generate artificial noise for the output of actor network, where $\sigma_{P}^{2}$ is annealing alongside the learning process to guide the TD3 agent from exploration to exploitation. Without loss of generality, the additive noise posed on the target actor network $\mathcal{N}^{-}$ is generated from zero-mean Normal distribution with annealing variance $\sigma_{P}^{2}$ as well. 

\begin{figure}[htbp]
\begin{algorithm}[H]
	\setstretch{1}
	\scriptsize
	\caption{The proposed hybrid D3QN-TD3 solution}\label{D3QNTD3algorithm}
	\SetKwData{Ini}{\bf{Initialization:}}
	{\Ini Initialize D3QN network $Q_{D3}(s,a\vert\boldsymbol{\theta}_{D3})$ and its target network $Q_{D3}(s,a\vert\boldsymbol{\theta}_{D3}^{-})$, with $\boldsymbol{\theta}_{D3}^{-}\leftarrow\boldsymbol{\theta}_{D3}$. Initialize TD3 network, including actor network $\mu(s\vert\boldsymbol{\theta}_{\mu})$, twin critics $Q_{P}(s,a\vert\boldsymbol{\theta}_{P_{\hat{j}}})$, target actor $\mu(s\vert\boldsymbol{\theta}_{\mu}^{-})$ and twin target critics $Q_{P}(s,a\vert\boldsymbol{\theta}_{P_{\hat{j}}}^{-})$, with $\boldsymbol{\theta}_{\mu}^{-}\leftarrow\boldsymbol{\theta}_{\mu}$ and $\boldsymbol{\theta}_{P_{\hat{j}}}^{-}\leftarrow\boldsymbol{\theta}_{P_{\hat{j}}}$. Initialize D3QN and TD3 replay buffers R and $\hat{\text{R}}$ with capacity $\grave{\text{D}}$ and $\acute{\text{D}}$, respectively\;\par}
	\For{$episode=[1,epi]$}{\par
		Initialize the outer environment and reset the UAV's location to $\vec{q}_{u}(0)$\;\par
		\For{$i=[1,epo_{outer}]$}{
			Observe the outer state $\mathbf{s}_{i}$\;
			Select the outer action $\mathbf{a}_{i}$, observe the available set  $\breve{\mathcal{B}}_{o}^{\mathbf{a}_{i}}$ and the occupied set  $\mathcal{B}_{o}^{\mathbf{a}_{i}}$\;
			Randomly select a BS $\breve{b}\in\breve{\mathcal{B}}_{o}^{\mathbf{a}_{i}}$ and check its {B2D} pathloss type, i.e., LoS or NLoS, then initialize the inner environment\;
			\For{$j=[1, epo_{inner}]$}{
				Observe the inner state $\hat{\mathbf{s}}_{j}$\;
				Select and execute the inner action $\hat{\mathbf{a}}_{j}$, observe the next inner state $\hat{\mathbf{s}}_{j + 1}$ and calculate the inner reward $\hat{\mathbf{r}}_{j}$\;
				Store transition $(\hat{\mathbf{s}}_{j}, \hat{\mathbf{a}}_{j}, \hat{\mathbf{s}}_{j + 1}, \hat{\mathbf{r}}_{j})$ into $\hat{\text{R}}$\;
				{\If{$card\left(\hat{\text{R}}\right)\ge N_{P}$}{
						Sample a mini-batch of $N_{P}$ transitions from $\hat{\text{R}}$, then update the twin critic networks $Q_{P}(s,a\vert\boldsymbol{\theta}_{P_{\hat{j}}})$ via gradient descent method in (\ref{descent_td3_critic})\;
						Every $N_{pud}$ times twin critics are trained, update actor network $\mu(s\vert\boldsymbol{\theta}_{\mu})$ via gradient ascent in (\ref{ascent_td3_actor}), and update target networks $Q_{P}(s,a\vert\boldsymbol{\theta}_{P_{\hat{j}}}^{-})$ and $\mu(s\vert\boldsymbol{\theta}_{\mu}^{-})$, following Polyak averaging rule in (\ref{Polyak_critic}) and (\ref{Polyak_actor}), respectively\;
				}}
				
			}
			
			Execute the outer action $\mathbf{a}_{i}$, observe the next outer state $\mathbf{s}_{i+1}$ and calculate the outer reward $\mathbf{r}_{i}$\;
			Store transition $(\mathbf{s}_{i}, \mathbf{a}_{i}, \mathbf{s}_{i + 1}, \mathbf{r}_{i})$ into R\;
			{\If{$card\left(\text{R}\right)\ge N_{D3}$}{
					Sample a mini-batch of $N_{D3}$ transitions from $\text{R}$, update D3QN network $Q_{D3}(s,a\vert\boldsymbol{\theta}_{D3})$ via gradient descent in (\ref{descent_d3qn})\;
					Update the D3QN target network $Q_{D3}(s,a\vert\boldsymbol{\theta}^{-}_{D3})$ every $\Upsilon_{D3}$ steps, i.e., $\boldsymbol{\theta}^{-}_{D3}\leftarrow\boldsymbol{\theta}_{D3}$\;
			}}
			
		}	
		Update $\epsilon\leftarrow\epsilon \times dec_{\epsilon}$ and $\sigma_{P}^{2}\leftarrow\sigma_{P}^{2}\times dec_{\sigma}$\;
	}	
\end{algorithm}

	\centering  
	\subfigcapskip=-.1cm
	\subfigure[Workflow diagram]{
		\label{Hybrid_D3QN_TD3_Diagram}
		\includegraphics[width=.45\linewidth]{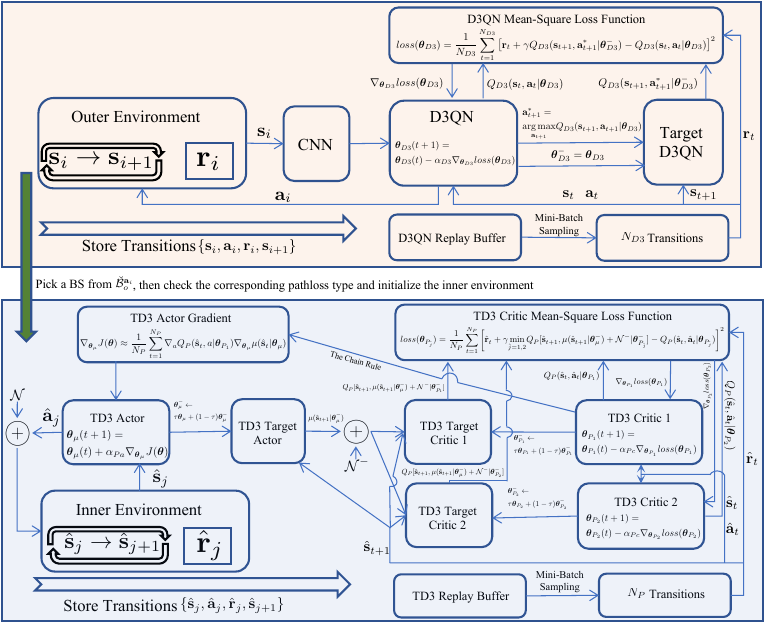}}
	\subfigure[{Offline exploitation}]{
		\label{An _instance_of_offline_exploitation_hybridd3qntd3}
		\includegraphics[width=.52\linewidth]{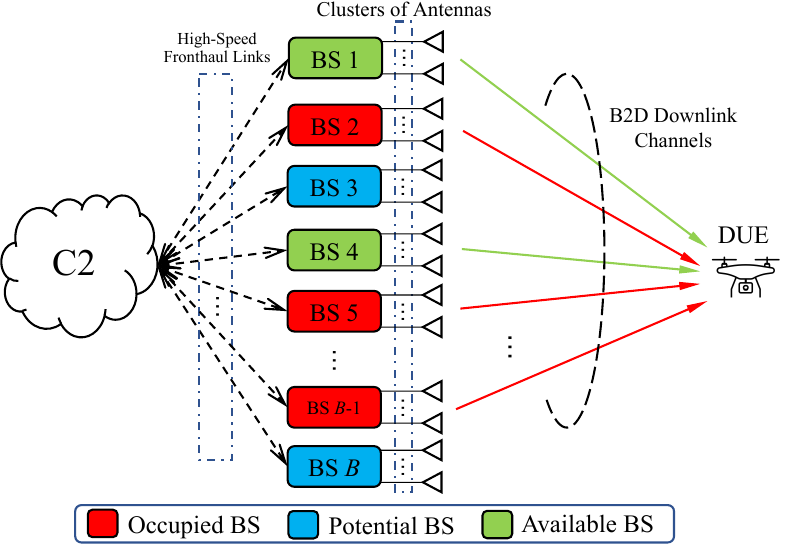}}		
	\caption{Workflow and an example illustrating offline exploitation of hybrid D3QN-TD3 solution}
	\label{Workflow_and_example_offline_exploitation}
	
\end{figure}

The overall pseudo-code and interacting diagram of the proposed hybrid D3QN-TD3 solution are given by \textbf{Algorithm} \textbf{\ref{D3QNTD3algorithm}} and Fig. \ref{Hybrid_D3QN_TD3_Diagram}, respectively. 

\subsubsection{Complexity Analysis and Justification of the Proposed D3QN-TD3 Algorithm}
{%
	The proposed D3QN-TD3 method is genuinely an online-centralized-learning-and-offline-decentralized-execution algorithm, for realizing its efficient implementation without introducing heavy burden of computations or unbearable delays and overheads of information transferring, e.g., imperfect CSIs and designed beamforming vectors between the C2 and available BSs, during its exploitation for RRM. Specifically, the proposed DRL-enabled algorithm is trained in a manner of online and centralized learning, aided by stochastic gradient descent/ascent with back-propagation, while interacting with outer and inner environments. Then, within the phase of offline exploitation, the trained D3QN agent would remain centralized at the C2, while the trained TD3 agent would be copied and distributed to be implanted to all the involved BSs, inspired by distributed ML frameworks, e.g., federated learning (FL) \cite{niknam2020federated, yang2020energy, amiri2020federated}. The C2 would select the optimal RB index for DUE after observing the current RBP map, with the help of trained D3QN agent. Furthermore, according to Subsection \ref{SubSec_RBallocation}, available BSs would be appointed by the C2, after which these available BSs would activate their TD3 agents to perform transmit beamforming. Therefore, D3QN and TD3 components carry out a two-step optimization process of RB coordination and beamforming design, via forward-propagations of the observed RBP map and local imperfect CSI, respectively. It is well known that, for feedforward neural networks, forward-propagation is much less computation-hungry than parameter update with back-propagation. For ease of interpreting, Fig. \ref{An _instance_of_offline_exploitation_hybridd3qntd3} shows an example of appointed available BSs, potential BSs, current occupied BSs and the corresponding B2D downlink transmissions, for C2's selected RB index. 
	
	During the online training phase, overheads rooted in interactions between the proposed D3QN-TD3 method and the environments as well as errors introduced by information observing and sharing would be another main source of concern. The outer reward function (\ref{reward_outer}) and inner reward function (\ref{inner_reward_function}) are designed to have nothing to do with extra environment information, e.g., CSIs and beamforming vectors of occupied BSs and AWGN variance, but focus on introducing more available BSs and optimizing beamforming performance of the selected available BS, respectively. These setups of reward functions significantly reduce overheads of information acquisitions during online training, while the aforementioned extra environment information is only used for calculating EOD for numerical results during offline execution. As the RBP map is genuinely a 2D binary matrix, it is assumed that the D3QN agent can observe it without errors or delays. Regarding the accuracy of available BS's CSI, estimation error has been modelled in (\ref{imperfectCSI}), for enhancing modelling practicality and highlighting the motivation of applying DRL-aided beamforming design. Thanks to the presence of experience replay buffer, both D3QN and TD3 are trained as per sampled mini-batch data out of their respective experience replay buffer, which means that their online trainings are off-policy and they could learn patterns of outer and inner environments from past experiences. Therefore, the TD3 agent is steered to select one single BS from current set of available BSs, for relieving issues of overheads and delays of CSI estimation and transfer during online learning stage. 
	Though each time TD3 agent interacts with only one available BS, with the help of experience replay buffer, TD3 agent can still be trained to learn patterns of the inner environment with sufficient amount of stored transitions.
}

\section{Simulation Results}\label{Sec_numerical_results}
\label{d3qn_ddpg_section_simulation_results}
{%
An urban subregion specified by $[0, 3]\times[0, 3]\times[0, 0.1]$ (in km) is concentrated for conducting numerical results, in which local building distribution is generated via one realization of ITU statistical model as shown in Fig. \ref{Building Distribution}. The parameter setting of this statistical model is in line with those in Subsection \ref{LoS_Check}. Note that the generated building distribution remains unchanged amid the entire simulation process, which corresponds to real-life scenario. In our considered model, the DUE's location at each time slot, i.e., $\vec{q}_{u}(n)$, is observed to determine the LoS/NLoS links via checking potential blockages between the DUE and the BSs.} 

{For ease of implementation and due to the trajectory-independent nature of formulated RRM problem (\ref{formulated_RRM_problem})}, the DUE's initial location and destination are fixed at $\vec{q}_{u}(I)\!=\!(1, 1, 0.1)$ km and $\vec{q}_{u}(D)\!=\!(2, 2, 0.1)$ km, respectively. The given trajectory is defined as the line between $\vec{q}_{u}(I)$ and $\vec{q}_{u}(D)$, of which the length is $\sqrt{\Vert\vec{q}_{u}(D)-\vec{q}_{u}(I)\Vert^{2}}\!\approx\!1.4$ km. Besides, the velocity of DUE is set as $V_{u}\!=\!35$ m/s and hence the DUE will spend $T_{u}\!=\!40$ s to travel between $\vec{q}_{u}(I)$ and $\vec{q}_{u}(D)$. Nakagami-{$m$} fading\footnote{In contrast to terrestrial communication scenarios where Rayleigh fading is widely applied to model small-scale fading, Rician or Nakagami-$m$ fading is more suitable to track the characteristics of {B2D} small-scale fading when LoS pathloss is experienced. For Nakagami-{$m$} fading model, special case $m=1$ is equivalent to Rayleigh fading while the case with $m>1$ can be utilized as an alternative of Rician fading where $m$ reflects the strength of LoS component.} is taken as an example to model the small-scale fading component for {B2D} channels in this paper. 
Besides, we apply the popularly-used Rayleigh fading \cite{li2020harvest} to model the terrestrial small-scale fading component and the beamforming strategy for terrestrial transmission is in line to $\vec{w}_{bg}=\vec{h}_{bg}^{\dagger}/\vert\vert\vec{h}_{bg}\vert\vert$ for simplicity.\footnote{This paper focuses on the interference management for cellular-connected UAV networks and the precoding configuration regarding terrestrial transmissions is not our interest. Here, we assume that the occupied BSs simply perform MRT technique for their serving GUEs.} Unless otherwise mentioned, the simulation parameter settings are in accordance with Table \ref{Simulation_Settings}.

\begin{table}[htbp]
	\centering
	\caption{Simulation parameter settings} 
	\label{Simulation_Settings}
	\tiny{}
	\begin{tabular}{p{3.4cm}|p{1.5cm}||p{3.2cm}|p{.75cm}||p{3.4cm}|p{1.1cm}} 
		\toprule
		\textbf{Parameters} & \textbf{Values}  & \textbf{Parameters} & \textbf{Values} & \textbf{Parameters} & \textbf{Values}\\ 
		\midrule
		
		Capacities of replay buffers $\grave{\text{D}}$/$\acute{\text{D}}$ & 100,000/100,000 & Given TOP threshold $\Gamma_{th}$ & 0 dB & Number of episodes $epi$ & 100\\
		
		Capacity of $\mathcal{B}$ & 37 & Number of outer epochs $epo_{outer}$ & 22 & Capacity of $\mathcal{K}$ & 100\\
		
		Number of inner epochs $epo_{inner}$ & 200 & Transmit power of each BS $P$ & 15 dBm & Target network update frequency $\Upsilon_{D3}$ & 500\\
		
		Number of antennas at each BS $M$ & 8 &Initial exploration parameter $\epsilon$/$\sigma_{P}^{2}$ & 0.9/1 & Tier of ICI $p$ & 1\\
		
		Exploration annealing rate $dec_{\epsilon}$/$dec_{\sigma}$ & 0.93/0.91 & Power of AWGN $\sigma^{2}$ & -90 dBm & Size of mini-batch $N_{D3}$/$N_{P}$ & 128/128 \\
		
		Carrier frequency $f_{c}$ & 2 GHz & Polyak interpolation factor $\tau$ & 0.00005 & DUE's altitude $h$/BS's antenna height $z$ & 100 m/25 m \\		
		
		Learning rates $\alpha_{D3}$/$\alpha_{Pc}$/$\alpha_{Pa}$ & 0.001/0.002/0.001 & SINR measurements $\varsigma$ & 1000 & Discount factor $\gamma$ & 0.99 \\ 
		
		Duration of time slot $\delta_{u}$ & 1.82 s & Nakagami factor $m$ for LoS/NLoS & 3/1  & Imperfect B2D CSI correlation factor $\rho$ & 0.75\\		
		
		Policy update delay factor $N_{pud}$  & 2& Prior-activation penalty coefficient $\kappa$ & 1 & Absolute saturation value of Tanh  $\xi$  & 2.5 \\
		
		Size of CNN's kernel $k_{1}^{c}$/$k_{2}^{c}$/$k_{3}^{c}$ & 5/4/3 & Number of CNN's filter  $f_{1}^{c}$/$f_{2}^{c}$/$f_{3}^{c}$  & 32/32/32& Size of CNN's stride $s_{1}^{c}$/$s_{2}^{c}$/$s_{3}^{c}$ & 1/1/1\\		

		ITU building distribution variable $\hat{\beta}$ & 103 buildings/km & Building distribution variables  $\hat{\alpha}$/$\hat{\gamma}$& 0.3/20 m&  Road width in building distribution $\hat{D}$ & 0.02 km\\				
		\bottomrule
	\end{tabular}
\end{table}

\subsection{Construction of DNNs}

{The proposed hybrid D3QN-TD3 solution is implemented on MacBook Pro with 2.3GHz quad-core Intel Core i5 and 8GB of 2133MHz LPDDR3 onboard memory, while the corresponding online training phase is performed on Python 3.8 with TensorFlow 2.3.1 and Keras.} The optimizer minimizing the mean-square error (MSE) for all the applied deep neural networks (DNNs) is \textit{Adam} with fixed learning rate. The activation function at each hidden layer (including each convolutional layer of CNN) is \textit{ReLU} function, for its simplicity and generality. Besides, the activation function utilized for both output layers in D3QN and critic network of TD3 is \textit{Linear}, while that for actor network of TD3 is \textit{Tanh}.\footnote{On the contrary to other popular activation functions, inter alia, \textit{ReLU}, \textit{Softmax} or \textit{Sigmoid}, \textit{Tanh} does not lose the degree of freedom to output both positive and negative values, which is of essence for the design of beamforming vector. Besides, the output of \textit{Tanh} is bounded within the range of (-1,1), which may enhance stability and robustness of training process.} 

The DNN of D3QN agent is constructed with fully connected feedforward ANN, in which 3 hidden layers contain 512, 256 and 128 neurons, respectively. The shapes of CNN's input and output layer of D3QN are determined by the dimension of RBP map and the number of possible RBs, i.e., $card(\mathcal{B})\times card(\mathcal{K})$ and $card(\mathcal{K})$, respectively. Before the output layer and after the last hidden layer, there is a duelling layer with $card(\mathcal{K}) + 1$ neurons, where one neuron reflects the estimation of state-value and the remaining $card(\mathcal{K})$ neurons track the action advantages for the $card(\mathcal{K})$ possible actions. After aggregation, the output layer generates the estimation of the $card(\mathcal{K})$ state-action values, as depicted in Fig. \ref{Duelling DQN Diagram}. 

Both the twin critic and actor networks' DNNs in TD3 agent are fully connected feedforward ANNs with 3 hidden layers consisting of 512, 256 and 128 neurons. The dimensions of input layer and output layer of the twin critic networks correspond to $2M + M + 2M$ and $1$, while those of the actor network are $2M + M$ and $2M$, respectively. This is because the Nakagami-$m$ fading component is in form of complex value, which should be decoupled at the input layers of the critic and actor networks. Besides, $M$ additional neurons should be added into the input layers of the critic and actor networks to help them identify LoS/NLoS inner environment. To calculate the inner reward function (\ref{inner_reward_function}), the actor network's outputs will be reconstructed into complex-value vector with $M\times1$ dimension, after which the vector will be normalized to satisfy constraint (\ref{Constraint22}).

Although activation function \textit{Tanh} is popular and effective, it may suffer from saturation. When the input of \textit{Tanh} locates in the left (right) saturation region, the corresponding output will unreasonably approach -1 (1), raising gradient vanishing issue amid back-propagation of the training process \cite{ding20203d}. To tackle this problem, prior-activation penalty will be posed onto the actor network's loss function, which can direct the input of \textit{Tanh} to remain in the range of unsaturation area. In implementation, gradient ascent on actor's expected return (\ref{ascent_td3_actor}) is accomplished via inverse batch gradient descent on the estimated Q function of critic 1 network, given by
\begin{equation}
	\boldsymbol{\theta}_{\mu}(t+1)=\boldsymbol{\theta}_{\mu}(t) - \alpha_{Pa}\nabla_{\boldsymbol{\theta}_{\mu}}loss(\boldsymbol{\theta}_{\mu}),
\end{equation}
where the mean loss function of actor network is denoted as
\begin{equation}
	loss(\boldsymbol{\theta}_{\mu}) = - \frac{1}{N_{P}}\sum_{t=1}^{N_{P}} Q_{P}\left[\hat{\mathbf{s}}_{t},\mu(\hat{\mathbf{s}}_{t}\vert\boldsymbol{\theta}_{\mu})\vert\boldsymbol{\theta}_{P_{1}}\right]. \label{mean_loss_function_actor_Chapter_hybrid_d3qntd3}
\end{equation}
Then, to perform prior-activation penalty trick, the mean loss function of actor network (\ref{mean_loss_function_actor_Chapter_hybrid_d3qntd3}) is rewritten as 
\begin{align}
	loss(\boldsymbol{\theta}_{\mu}) = &\frac{1}{N_{P}}\sum_{t=1}^{N_{P}} \left\{- Q_{P}\left[\hat{\mathbf{s}}_{t},\mu(\hat{\mathbf{s}}_{t}\vert\boldsymbol{\theta}_{\mu})\vert\boldsymbol{\theta}_{P_{1}}\right] + \right.\nonumber\\
	&\left.\kappa\left[\max\left(\frac{1}{2M}\sum_{m=1}^{2M}\varrho_{t,m} - \xi, 0\right) + \max\left(-\frac{1}{2M}\sum_{m=1}^{2M}\varrho_{t,m} - \xi, 0\right)\right]^{2}\right\},\label{prior_activation_mean_loss_function_actor_Chapter_hybrid_d3qntd3}
\end{align}
where $\kappa$ indicates the coefficient of prior-activation penalty, $\xi$ represents the absolute saturation value of \textit{Tanh} activation function, and $\varrho_{t,m}$ denotes the prior-activation value of the corresponding neuron $m=\{1,2,\cdots,2M\}$ over one time of sampling $t$ from mini-batch transitions. The actor is trained to minimize (\ref{prior_activation_mean_loss_function_actor_Chapter_hybrid_d3qntd3}), which can directly navigate the prior-activation values of actor's output neurons to remain in the unsaturation region and thus helping circumvent the issue of gradient vanishing caused by saturation.

\begin{figure}[htbp]
	\centering  
	\subfigcapskip=-.2cm
	\subfigure[Reward history of D3QN]{
		\label{reward_history_d3qn}
		\includegraphics[scale=0.525]{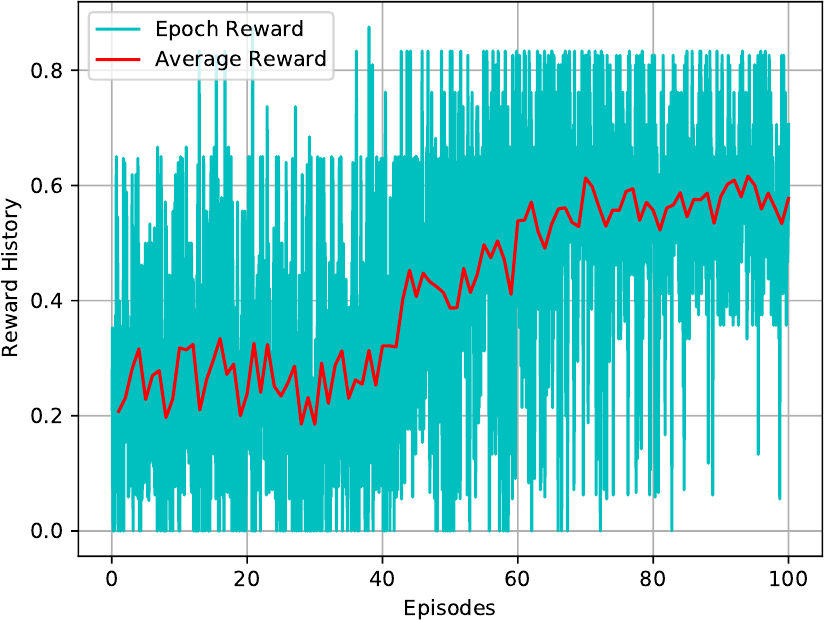}}
	\hspace{1cm}
	\subfigure[Reward history of TD3]{
		\label{reward_history_td3}
		\includegraphics[scale=0.525]{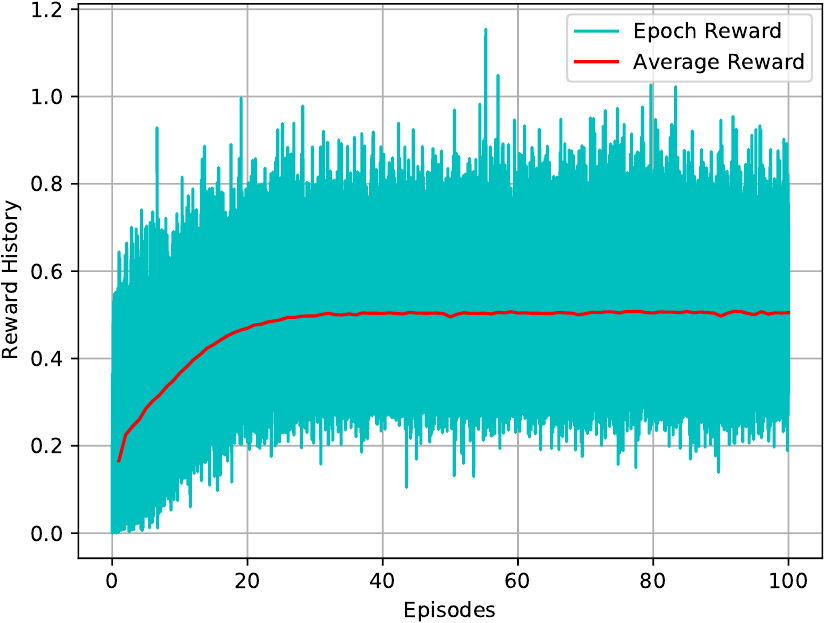}}
	\caption{Reward history }
	\label{Reward_history}
\end{figure}
\subsection{Training of Hybrid D3QN-TD3 Algorithm}

Fig. \ref{Reward_history} shows reward history curves versus training episodes for the proposed hybrid D3QN-TD3 solution. The average reward reflects the expected value of epoch rewards for each episode, which is calculated via averaging accumulated rewards over training epochs. It can be observed from Fig. \ref{Reward_history} that both D3QN and TD3 networks illustrate increasing trending of average reward alongside the training process, though experiencing some fluctuations that are usual phenomena in the regime of DRL-related algorithms. Specifically, the D3QN's average reward converges to the optimum (around 0.57) after 70 training episodes, while the TD3 converges to its highest average reward (about 0.51) after 40 training episodes. Fig. \ref{reward_history_d3qn} validates that the D3QN agent can adapt to the dynamic RBP environment via allocating proper RB index to the DUE for each time slot, while Fig. \ref{reward_history_td3} verifies that the TD3 agent is able to adjust transmit beamforming vectors to fit the small-scale fading environment. After saving the hybrid D3QN-TD3 model with the highest average rewards, it can be re-loaded to realize EOD performance comparison which will be illustrated in Subsection \ref{subsection_performance_comparison}.

\begin{figure}[htbp]
	\centering  
	\subfigcapskip=-.2cm
	\subfigure[Impact of $\alpha_{D3}$]{
		\label{curves_various_learning_rate_d3qn}
		\includegraphics[scale=0.5]{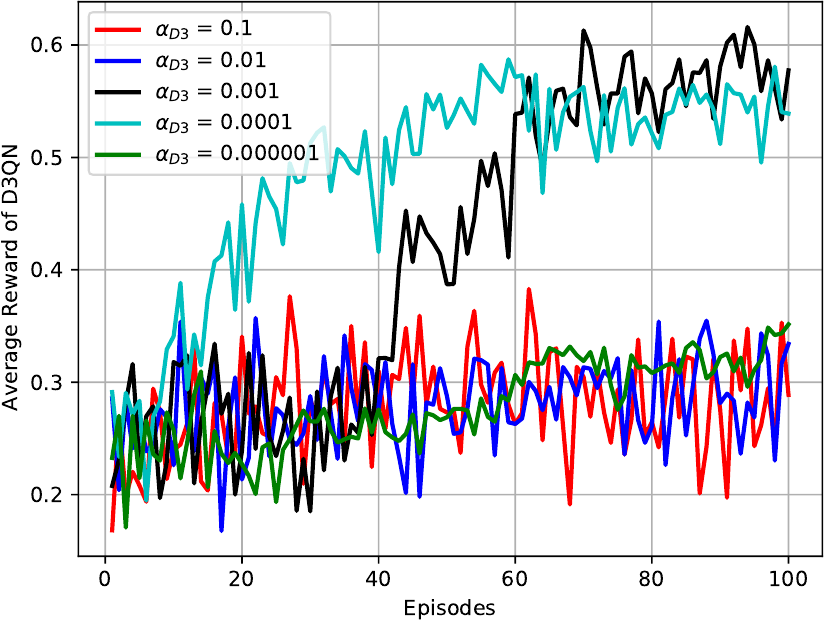}}
	\hspace{1cm}
	\subfigure[Impact of $\Upsilon_{D3}$]{
		\label{curves_various_target_updates_d3qn}
		\includegraphics[scale=0.5]{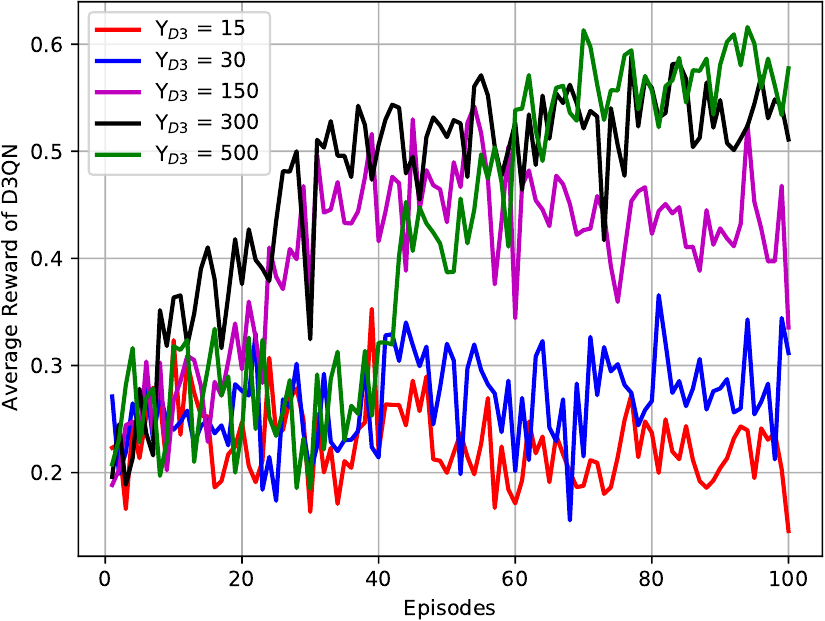}}
	\caption{Impact of learning rates and target network update frequency}
	\label{Impact_of_learning_rates_and_target_update_frequency}
\end{figure}
\begin{figure}[htbp]
	\centering  
	\subfigcapskip=-.2cm
	\subfigure[Impact of $\alpha_{Pa}$ and $\alpha_{Pc}$]{
		\label{curves_various_learning_rate_td3}
		\includegraphics[scale=0.5]{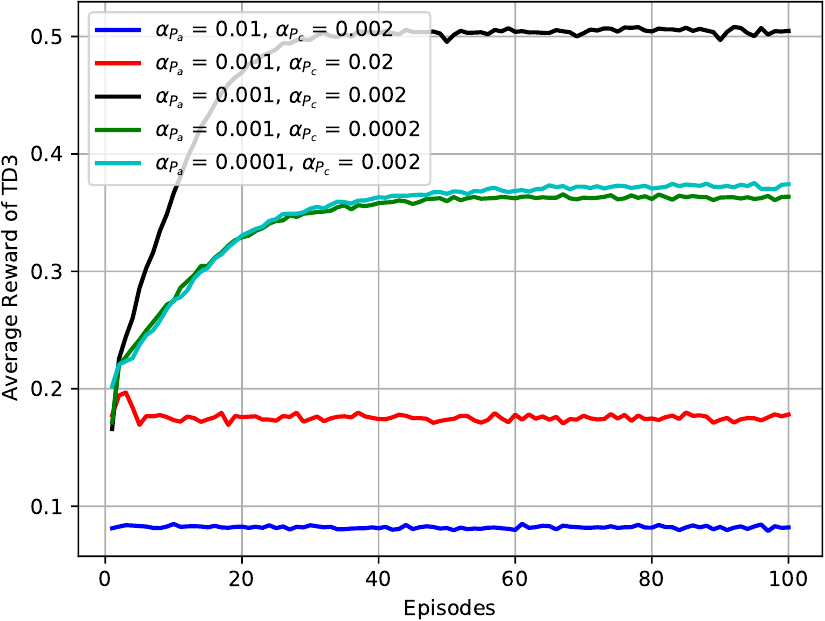}}
	\hspace{1cm}
	\subfigure[Impact of $\tau$]{
		\label{curves_various_target_soft_updates_td3}
		\includegraphics[scale=0.5]{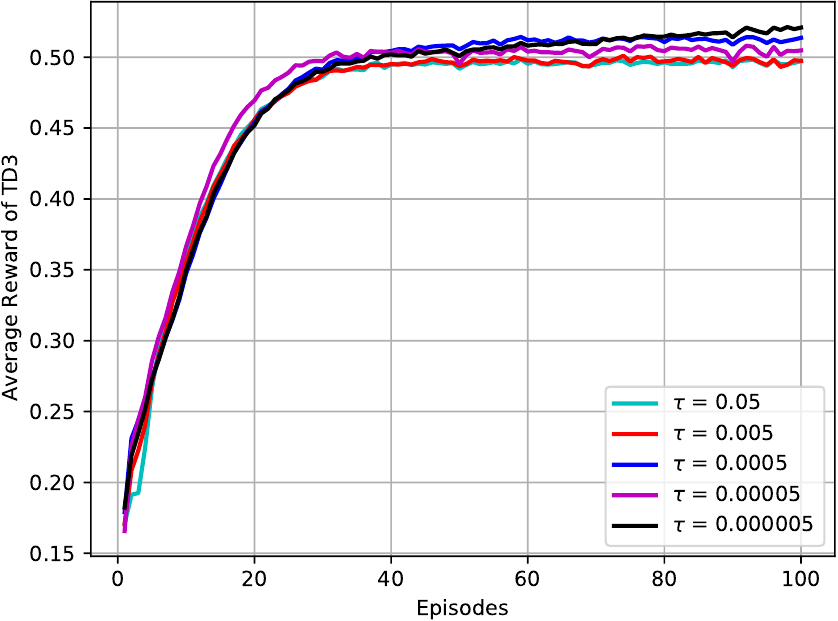}}
	\caption{Impact of learning rates and Polyak interpolation factor}
	\label{Impact_of_learning_rates_and_Polyak_interpolation_factor}
\end{figure}
\subsection{Impacts of Hyper-parameters}

It is well known that the overall performance of DRL-related algorithms is sensitive to hyper-parameters, e.g., target network update and learning rate. The hyper-parameters should be picked carefully for given system settings, to realize satisfactory learning quality and convergence speed.  

Fig. \ref{curves_various_learning_rate_d3qn} delivers average D3QN reward curves versus training episodes with various $\alpha_{D3}$, while Fig. \ref{curves_various_learning_rate_td3} demonstrates average TD3 reward curves versus training episodes with different combinations of $\alpha_{Pa}$ and $\alpha_{Pc}$. From these figures, it can be observed that learning rates pose significant impacts on learning performance and convergence speed. With relatively high $\alpha_{D3}$, i.e., $\alpha_{D3}=\{0.1, 0.01\}$, although the D3QN's convergences are quite rapid, it reaches extremely unsatisfactory learning scores (both around 0.3). With relatively small $\alpha_{D3}$, i.e., $\alpha_{D3}=\{0.001, 0.0001\}$, the D3QN agent can achieve higher scores (about 0.57 and 0.54, respectively). Surprisingly, when $\alpha_{D3}$ is extremely small, i.e., $\alpha_{D3}=0.000001$, it leads to unsatisfactory learning performance in the range of 100 training episodes. However, $\alpha_{D3}=0.000001$ may have the potential to help the D3QN agent reach a new highest score, for which the price is that much more training episodes are needed (i.e., less favourable convergence rate). For Fig. \ref{curves_various_learning_rate_td3}, learning rate combination $[\alpha_{Pa}=0.001,\alpha_{Pc}=0.002]$ is selected as the anchor for comparison, which can converge to its optimal score (around 0.51) after about 40 training episodes. With higher $\alpha_{Pa}$, i.e., $[\alpha_{Pa}=0.01,\alpha_{Pc}=0.002]$, the TD3 agent barely learns anything and achieves significantly worse score (around 0.06). 
With smaller $\alpha_{Pa}$, i.e., $[\alpha_{Pa}=0.0001,\alpha_{Pc}=0.002]$, the TD3 agent converges to a worse score (about 0.38) than the anchor, after around 60 training episodes, which means that it experiences slower convergence rate. 
With higher $\alpha_{Pc}$, i.e., $[\alpha_{Pa}=0.001,\alpha_{Pc}=0.02]$, the TD3 agent converges to worse learning quality (around 0.18), although the corresponding convergence speed is relatively rapid. 
With smaller $\alpha_{Pc}$, i.e., $[\alpha_{Pa}=0.001,\alpha_{Pc}=0.0002]$, the TD3 agent can only reach much lower learning score (around 0.37), while experiencing a comparable convergence speed (converging after around 40 training episodes). 
From the above observations, it is straightforward to conclude that the proposed hybrid D3QN-TD3 solution is unsurprisingly sensitive to learning rate which should be selected delicately for accomplishing a good trade-off between learning quality and convergence speed. 

Fig. \ref{curves_various_target_updates_d3qn} depicts average D3QN reward curves versus training episodes with different $\Upsilon_{D3}$, while Fig. \ref{curves_various_target_soft_updates_td3} illustrates average TD3 reward curves versus training episodes with various $\tau$. From these figures, it can be easily concluded that target network technique adopted in the proposed hybrid D3QN-TD3 algorithm is undoubtedly of essence. Specifically, less frequent updating (i.e., larger $\Upsilon_{D3}$) on D3QN's target network can help the D3QN agent achieve better learning scores, while less amount of updating (i.e., smaller $\tau$) on TD3's target networks is more favourable. However, larger $\Upsilon_{D3}$ and smaller $\tau$ may result in slower convergence speed. Hence, the picking of $\Upsilon_{D3}$ and $\tau$ is important for the proposed hybrid D3QN-TD3 solution to deal with the dilemma between learning performance and convergence speed.

\begin{figure}[htbp]
	\centering  
	\subfigcapskip=-.2cm
	\subfigure[Performance comparison versus $P$]{
		\label{performance_comparison_P}
		\includegraphics[scale=0.5]{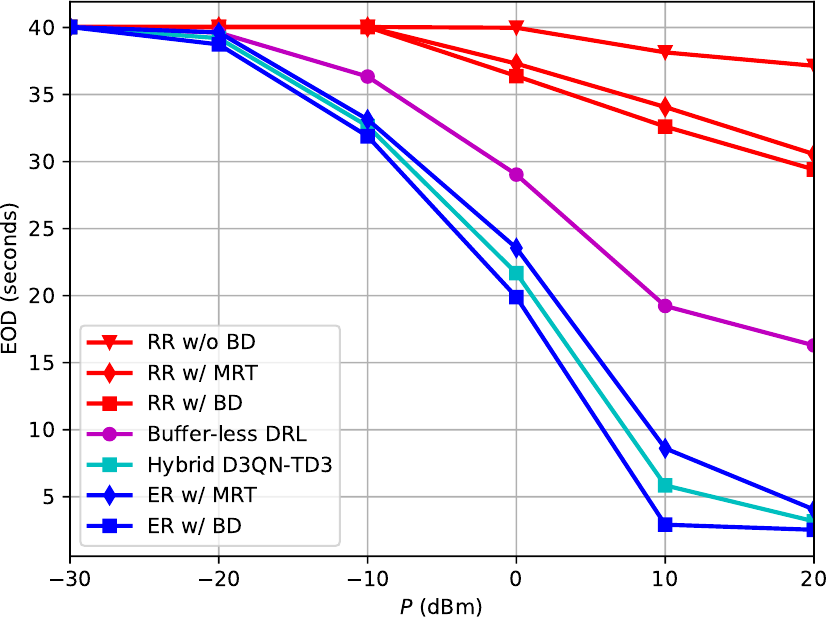}}
	\hspace{1cm}
	\subfigure[Performance comparison versus $M$]{
		\label{performance_comparison_M}
		\includegraphics[scale=0.505]{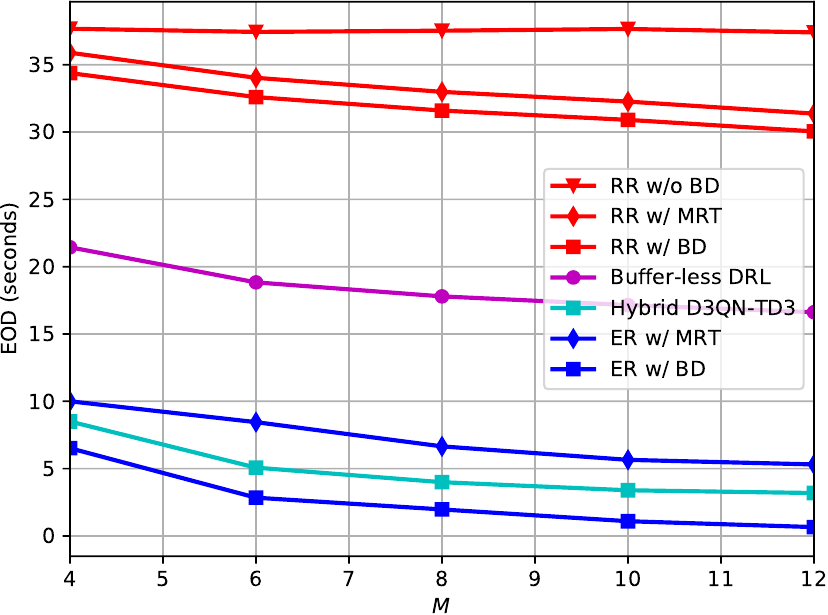}}
	\caption{Performance comparison}
	\label{performance_comparison}
\end{figure}

\subsection{Performance Comparison}
\label{subsection_performance_comparison}

{After online centralized training, performance comparison between representative baselines and the trained D3QN-TD3 solution can be conducted within the offline decentralized exploitation phase, where} the following benchmarks are provided. 1) \textit{RR w/o BD}: the RB index selected for each time slot and the beamforming vector at each available BS are both randomly generated. Note that this approach is supposed to be the worst, which may lead the DUE to suffer from the maximal transmission outage duration.
2) \textit{RR w/ BD}: the RB index scheduled for each time slot is randomly selected, but the beamforming vectors at available BSs are generated with the help of trained TD3 agent. 3) \textit{RR w/ MRT}: different from \textit{RR w/ BD}, MRT technique is invoked to generate the beamforming vectors, based on the corresponding estimated CSIs. 4) \textit{ER w/ BD}: the RB index assigned for each time slot is the optimal via exhaustive search method, which can maximize (\ref{reward_outer}) for every observed RBP map. Besides, the beamforming vector at each available BS is obtained from the trained TD3 agent. Note that this benchmark serves as the lower bound of EOD performance, which is supposed to help the DUE suffer the minimal transmission outage duration. 5) \textit{ER w/ MRT}: different from \textit{ER w/ BD}, the beamforming vectors are designed with the help of MRT technique, based on the corresponding estimated CSIs. {6) \textit{Buffer-less DRL}: this DRL-aided baseline is invoked to highlight the importance of experience replay buffer, to construct which state-of-the-art synchronous advantage actor-critic (A2C) \cite{mnih2016asynchronous} and proximal policy gradient (PPO) \cite{schulman2017proximal} frameworks are invoked. During offline exploitation for RRM, RB index selection and beamforming design are alternatively accomplished via the trained A2C and PPO agents, respectively. Please note that extra hyper-parameters introduced by A2C and PPO, e.g., value function coefficient, entropy parameter, the maximum value for gradient clipping and policy-clipping ratio controlling the distance between new policy and its old counterpart, are set in line with those default values as suggested by \cite{stable-baselines3}.}

The proposed hybrid D3QN-TD3 solution provides the proper RB index for each time slot and designed beamforming vector for each available BS, with the aid of trained D3QN agent and TD3 agent, respectively. Fig. \ref{performance_comparison_P} and Fig. \ref{performance_comparison_M} show EOD curves of the proposed D3QN-TD3 solution and benchmarks versus $P$ and $M$, respectively. It is clearly illustrated in Fig. \ref{performance_comparison_P} that the EOD curves decrease dramatically with the increase of $P$, which means that higher $P$ can help the DUE achieve better transmission outage performance (i.e., lower EOD). Comparing the EOD curves of \textit{RR w/o BD} and \textit{RR w/ BD}, EOD performance enhancement can be observed (especially, for $P\in[-10,20]$ dBm), which validates the effectiveness of TD3 component. Furthermore, via comparing the curves of \textit{RR w/ BD} and \textit{RR w/ MRT}, one can observe that the trained TD3 agent can help the UAV suffer from less amount of EOD than MRT beamforming scheme (for $P\in[-10,20]$ dBm), in case of imperfect CSI estimation. Similar phenomenon can be observed via comparing \textit{ER w/ BD} and \textit{ER w/ MRT}. This is because the MRT beamforming strategy can only adapt to the estimated CSI, while the TD3 agent is trained to adapt to the overall imperfect CSI. Besides, greater EOD performance improvement can be achieved with the help of D3QN component, via comparing the EOD curves of \textit{RR w/ BD} and the proposed hybrid D3QN-TD3 solution (especially, for $P\in[-20, 20]$ dBm). The aforementioned observations validate that the {trained} D3QN and TD3 agents are able to offer independent EOD performance gains, which is a remarkable feature of the proposed hybrid D3QN-TD3 solution. Compared to the optimal method \textit{ER w/ BD}, the proposed hybrid D3QN-TD3 solution can help the DUE achieve suboptimal EOD performance which performs slightly worse than the optimal approach but can provide significant EOD reduction than benchmarks \textit{RR w/o BD}, \textit{RR w/ BD} and \textit{RR w/ MRT}. Most importantly, the proposed hybrid D3QN-TD3 solution outperforms \textit{ER w/ MRT} as well, which means that the joint RB allocation and beamforming design provided by the proposed hybrid D3QN-TD3 solution can offer more significant EOD reduction than that offered by MRT beamforming with optimal RB allocation. {Comparing curves of \textit{Buffer-less DRL} and the proposed hybrid D3QN-TD3 algorithm, it is straightforward to find that the latter outperforms the former counterpart with significant performance gap. This phenomenon reflects the importance of experience replay buffer for DRL-aided solution solving the concentrated EOD minimization problem, because the experience replay buffer makes it possible to break the temporal correlations of experienced transitions via mixing recent and former experiences into the replay buffer, guaranteeing that rarely experienced transitions get fairer chances to be utilized. Through scarifying computation and memory for recording and sampling, experience replay technique lightens  burden of requiring numerous experiences for training. However, this compromise is worthwhile because the interactions between RL agent and environment are more resource-expensive in general \cite{schaul2016prioritized}.} Similar conclusions can be drawn from Fig. \ref{performance_comparison_M} which demonstrates EOD curves versus various $M$. Note that for specific antenna number configuration, the proposed hybrid D3QN-TD3 algorithm needs to be retrained with the corresponding antenna number.\footnote{The robustness of TD3 agent to various antenna number configurations can be further enhanced via adopting hypernetwork \cite{goutay2020deep}, which is envisioned to relieve the retraining burden or even liberate the TD3 agent from being retained.} From this figure, one can find the other fact that increasing $M$ can help enhance EOD performance for solutions with beamforming design (\textit{RR w/ BD}, \textit{RR w/ MRT}, \textit{Buffer-less DRL}, \textit{Hybrid D3QN-TD3}, \textit{ER w/ BD} and \textit{ER w/ MRT}), but cannot achieve any EOD reduction for solution without beamforming design (\textit{RR w/o BD}).

\section{Conclusion}\label{d3qn_ddpg_section_conclusion}

{This paper studied an RRM problem of joint RB allocation and beamforming design for cellular-connected UAVs while protecting B2G transmissions from being interfered by co-channel ICIs generated from B2D communications, in which DUE's EOD was minimized via the proposed DRL-aided hybrid D3QN-TD3 algorithm.} Specifically, the D3QN and TD3 agents were trained to accomplish RB coordination in discrete action domain and beamforming design in continuous action regime, respectively. To realize this, an outer MDP was defined to characterize the dynamic RBP environment at the terrestrial BSs, while the inner MDP was formulated to trace the time-varying feature of B2D small-scale fading. The hybrid D3QN-TD3 solution was proposed to solve the outer MDP and the inner MDP interactively so that suboptimal EOD performance for the considered optimization problem can be achieved. {Numerical results illustrated that the proposed hybrid D3QN-TD3 algorithm can significantly reduce EOD suffered by the DUE compared to the provided benchmarks, where the trained D3QN and TD3 agents were also validated to offer independent improvements on EOD minimization performance.}

\bibliographystyle{IEEEtran}
{\bibliography{reference}}

\begin{thebibliography}{10}
\providecommand{\url}[1]{#1}
\csname url@samestyle\endcsname
\providecommand{\newblock}{\relax}
\providecommand{\bibinfo}[2]{#2}
\providecommand{\BIBentrySTDinterwordspacing}{\spaceskip=0pt\relax}
\providecommand{\BIBentryALTinterwordstretchfactor}{4}
\providecommand{\BIBentryALTinterwordspacing}{\spaceskip=\fontdimen2\font plus
\BIBentryALTinterwordstretchfactor\fontdimen3\font minus
  \fontdimen4\font\relax}
\providecommand{\BIBforeignlanguage}[2]{{%
\expandafter\ifx\csname l@#1\endcsname\relax
\typeout{** WARNING: IEEEtran.bst: No hyphenation pattern has been}%
\typeout{** loaded for the language `#1'. Using the pattern for}%
\typeout{** the default language instead.}%
\else
\language=\csname l@#1\endcsname
\fi
#2}}
\providecommand{\BIBdecl}{\relax}
\BIBdecl

\bibitem{alsamhi2021green}
S.~H. Alsamhi, F.~Afghah, R.~Sahal, A.~Hawbani, M.~A. Al-qaness, B.~Lee, and
  M.~Guizani, ``Green internet of things using {UAV}s in {B5G} networks: A
  review of applications and strategies,'' \emph{Elsevier Ad Hoc Netw.}, vol.
  117, p. 102505, Jun. 2021.

\bibitem{wang2022uav}
S.~Wang, S.~Hosseinalipour, M.~Gorlatova, C.~G. Brinton, and M.~Chiang,
  ``{UAV}-assisted online machine learning over multi-tiered networks: A
  hierarchical nested personalized federated learning approach,'' \emph{IEEE
  Trans. Netw. Service Manag.}, Oct. 2022.

\bibitem{li2021energy}
H.~Li, J.~Li, M.~Liu, Z.~Ding, and F.~Gong, ``Energy harvesting and resource
  allocation for cache-enabled {UAV} based {IoT NOMA} networks,'' \emph{IEEE
  Trans. Veh. Technol.}, vol.~70, no.~9, pp. 9625--9630, Jul. 2021.

\bibitem{liu2019multi}
L.~Liu, S.~Zhang, and R.~Zhang, ``Multi-beam {UAV} communication in cellular
  uplink: Cooperative interference cancellation and sum-rate maximization,''
  \emph{IEEE Trans. Wireless Commun.}, vol.~18, no.~10, pp. 4679--4691, Jul.
  2019.

\bibitem{zeng2019path}
Y.~Zeng and X.~Xu, ``Path design for cellular-connected {UAV} with
  reinforcement learning,'' in \emph{Proc. IEEE Global Commun. Conf.
  (GLOBECOM)}, Waikoloa, Dec. 2019, pp. 1--6.

\bibitem{zeng2021simultaneous}
Y.~Zeng, X.~Xu, S.~Jin, and R.~Zhang, ``Simultaneous navigation and radio
  mapping for cellular-connected {UAV} with deep reinforcement learning,''
  \emph{IEEE Trans. Wireless Commun.}, vol.~20, no.~7, pp. 4205--4220, Feb.
  2021.

\bibitem{li2022path}
Y.~Li, A.~H. Aghvami, and D.~Dong, ``Path planning for cellular-connected
  {UAV}: A {DRL} solution with quantum-inspired experience replay,'' \emph{IEEE
  Trans. Wireless Commun.}, vol.~21, no.~10, pp. 7897--7912, Apr. 2022.

\bibitem{zhang2018cellular}
S.~Zhang, Y.~Zeng, and R.~Zhang, ``Cellular-enabled {UAV} communication: A
  connectivity-constrained trajectory optimization perspective,'' \emph{IEEE
  Trans. Commun.}, vol.~67, no.~3, pp. 2580--2604, Nov. 2018.

\bibitem{wu2021comprehensive}
Q.~Wu, J.~Xu, Y.~Zeng, D.~W.~K. Ng, N.~Al-Dhahir, R.~Schober, and A.~L.
  Swindlehurst, ``A comprehensive overview on {5G}-and-beyond networks with
  {UAV}s: From communications to sensing and intelligence,'' \emph{IEEE J. Sel.
  Areas Commun.}, vol.~39, no.~10, pp. 2912--2945, Jun. 2021.

\bibitem{chandhar2017massive}
P.~Chandhar, D.~Danev, and E.~G. Larsson, ``Massive {MIMO} for communications
  with drone swarms,'' \emph{IEEE Trans. Wireless Commun.}, vol.~17, no.~3, pp.
  1604--1629, Dec. 2017.

\bibitem{senadhira2020uplink}
N.~Senadhira, S.~Durrani, X.~Zhou, N.~Yang, and M.~Ding, ``Uplink {NOMA} for
  cellular-connected {UAV}: Impact of {UAV} trajectories and altitude,''
  \emph{IEEE Trans. Commun.}, vol.~68, no.~8, pp. 5242--5258, May 2020.

\bibitem{mei2019cellular}
W.~Mei, Q.~Wu, and R.~Zhang, ``Cellular-connected {UAV}: Uplink association,
  power control and interference coordination,'' \emph{IEEE Trans. Wireless
  Commun.}, vol.~18, no.~11, pp. 5380--5393, Aug. 2019.

\bibitem{pang2019uplink}
X.~Pang, G.~Gui, N.~Zhao, W.~Zhang, Y.~Chen, Z.~Ding, and F.~Adachi, ``Uplink
  precoding optimization for {NOMA} cellular-connected {UAV} networks,''
  \emph{IEEE Trans. Commun.}, vol.~68, no.~2, pp. 1271--1283, Nov. 2019.

\bibitem{mei2019cooperative}
W.~Mei and R.~Zhang, ``Cooperative downlink interference transmission and
  cancellation for cellular-connected {UAV}: A divide-and-conquer approach,''
  \emph{IEEE Trans. Commun.}, vol.~68, no.~2, pp. 1297--1311, Nov. 2019.

\bibitem{hattab2020energy}
G.~Hattab and D.~Cabric, ``Energy-efficient massive {IoT} shared spectrum
  access over {UAV}-enabled cellular networks,'' \emph{IEEE Trans. Commun.},
  vol.~68, no.~9, pp. 5633--5648, May 2020.

\bibitem{li2022intelligent}
Y.~Li and A.~H. Aghvami, ``Intelligent {UAV} navigation: A {DRL-QiER}
  solution,'' in \emph{Proc. IEEE Int. Conf. Commun. (ICC)}, Seoul, May 2022,
  pp. 1--6.

\bibitem{zhou2018uav}
F.~Zhou, Y.~Wu, H.~Sun, and Z.~Chu, ``{UAV}-enabled mobile edge computing:
  Offloading optimization and trajectory design,'' in \emph{Proc. IEEE Int.
  Conf. Commun. (ICC)}, Kansas, USA, May 2018, pp. 1--6.

\bibitem{zhou2018mobile}
Z.~Zhou, J.~Feng, B.~Gu, B.~Ai, S.~Mumtaz, J.~Rodriguez, and M.~Guizani, ``When
  mobile crowd sensing meets {UAV}: Energy-efficient task assignment and route
  planning,'' \emph{IEEE Trans. Commun.}, vol.~66, no.~11, pp. 5526--5538, Jul.
  2018.

\bibitem{chu2019uav}
Z.~Chu, W.~Hao, P.~Xiao, and J.~Shi, ``{UAV} assisted spectrum sharing
  ultra-reliable and low-latency communications,'' in \emph{Proc. IEEE Global
  Commun. Conf. (GLOBECOM)}, Waikoloa, USA, Dec. 2019, pp. 1--6.

\bibitem{hu2019optimal}
J.~Hu, Y.~Wu, R.~Chen, F.~Shu, and J.~Wang, ``Optimal detection of {UAV}'s
  transmission with beam sweeping in covert wireless networks,'' \emph{IEEE
  Trans. Veh. Technol.}, vol.~69, no.~1, pp. 1080--1085, Oct. 2019.

\bibitem{wang2020energy}
W.~Wang, X.~Li, M.~Zhang, K.~Cumanan, D.~W.~K. Ng, G.~Zhang, J.~Tang, and O.~A.
  Dobre, ``Energy-constrained {UAV}-assisted secure communications with
  position optimization and cooperative jamming,'' \emph{IEEE Trans. Commun.},
  vol.~68, no.~7, pp. 4476--4489, Apr. 2020.

\bibitem{boudreau2009interference}
G.~Boudreau, J.~Panicker, N.~Guo, R.~Chang, N.~Wang, and S.~Vrzic,
  ``Interference coordination and cancellation for 4{G} networks,'' \emph{IEEE
  Commun. Mag.}, vol.~47, no.~4, pp. 74--81, May 2009.

\bibitem{kosta2012interference}
C.~Kosta, B.~Hunt, A.~U. Quddus, and R.~Tafazolli, ``On interference avoidance
  through inter-cell interference coordination ({ICIC}) based on {OFDMA} mobile
  systems,'' \emph{IEEE Commun. Surveys Tuts.}, vol.~15, no.~3, pp. 973--995,
  Dec. 2012.

\bibitem{zhang2010dynamic}
R.~Zhang, Y.-C. Liang, and S.~Cui, ``Dynamic resource allocation in cognitive
  radio networks,'' \emph{IEEE Signal Process. Mag.}, vol.~27, no.~3, pp.
  102--114, Apr. 2010.

\bibitem{irmer2011coordinated}
R.~Irmer \emph{et~al.}, ``Coordinated multipoint: Concepts, performance, and
  field trial results,'' \emph{IEEE Commun. Mag.}, vol.~49, no.~2, pp.
  102--111, Feb. 2011.

\bibitem{series2013propagation}
P.~Series, ``Propagation data and prediction methods required for the design of
  terrestrial broadband radio access systems operating in a frequency range
  from 3 to 60 {GH}z,'' \emph{Recommendation ITU-R}, pp. 1410--1415, 2013.

\bibitem{wang2021jamming}
X.~Wang, M.~C. Gursoy, T.~Erpek, and Y.~E. Sagduyu, ``Jamming-resilient path
  planning for multiple {UAV}s via deep reinforcement learning,'' in
  \emph{Proc. IEEE Int. Conf. Commun. Workshops (ICC Workshops)}, Montreal,
  Jun. 2021, pp. 1--6.

\bibitem{ding20203d}
R.~Ding, F.~Gao, and X.~S. Shen, ``{3D UAV} trajectory design and frequency
  band allocation for energy-efficient and fair communication: A deep
  reinforcement learning approach,'' \emph{IEEE Trans. Wireless Commun.},
  vol.~19, no.~12, pp. 7796--7809, Aug. 2020.

\bibitem{ThreeGPP2017}
{\relax 3GPP TR 36.777}, ``Enhanced {LTE} support for aerial vehicles,'' Dec.
  2017.

\bibitem{li2022covertness}
Y.~Li and A.~H. Aghvami, ``Covertness-aware trajectory design for {UAV}: A
  multi-step {TD3-PER} solution,'' in \emph{Proc. IEEE Int. Conf. Commun.
  (ICC)}, Seoul, May 2022, pp. 1--6.

\bibitem{mudumbai2009distributed}
R.~Mudumbai, D.~R.~B. Iii, U.~Madhow, and H.~V. Poor, ``Distributed transmit
  beamforming: challenges and recent progress,'' \emph{IEEE Commun. Mag.},
  vol.~47, no.~2, pp. 102--110, Feb. 2009.

\bibitem{joudeh2016sum}
H.~Joudeh and B.~Clerckx, ``Sum-rate maximization for linearly precoded
  downlink multiuser {MISO} systems with partial {CSIT}: A rate-splitting
  approach,'' \emph{IEEE Trans. Commun.}, vol.~64, no.~11, pp. 4847--4861, Aug.
  2016.

\bibitem{choi2017joint}
J.~Choi, ``Joint rate and power allocation for {NOMA} with statistical {CSI},''
  \emph{IEEE Trans. Commun.}, vol.~65, no.~10, pp. 4519--4528, Jun. 2017.

\bibitem{xiao2021uav}
L.~Xiao, Y.~Ding, J.~Huang, S.~Liu, Y.~Tang, and H.~Dai, ``{UAV} anti-jamming
  video transmissions with {QoE} guarantee: A reinforcement learning-based
  approach,'' \emph{IEEE Trans. Commun.}, vol.~69, no.~9, pp. 5933--5947, Jun.
  2021.

\bibitem{kadan2021theoretical}
F.~E. Kadan and A.~{\"O}. Y{\i}lmaz, ``A theoretical performance bound for
  joint beamformer design of wireless fronthaul and access links in downlink
  {C-RAN},'' \emph{IEEE Trans. Wireless Commun.}, vol.~21, no.~4, pp.
  2177--2192, Sep. 2021.

\bibitem{elhoushy2021cell}
S.~Elhoushy, M.~Ibrahim, and W.~Hamouda, ``Cell-free massive {MIMO}: A
  survey,'' \emph{IEEE Commun. Surveys Tuts.}, vol.~24, no.~1, pp. 492--523,
  Oct. 2021.

\bibitem{wang2016dueling}
Z.~Wang, T.~Schaul, M.~Hessel, H.~Hasselt, M.~Lanctot, and N.~Freitas,
  ``Dueling network architectures for deep reinforcement learning,'' in
  \emph{Proc. Int. Conf. Mach. Learn. (ICML)}, Jun. 2016, pp. 1995--2003.

\bibitem{lillicrap2015continuous}
T.~P. Lillicrap, J.~J. Hunt, A.~Pritzel, N.~Heess, T.~Erez, Y.~Tassa,
  D.~Silver, and D.~Wierstra, ``Continuous control with deep reinforcement
  learning,'' in \emph{Proc. Int. Conf. Learn. Represent. (ICLR)}, San Juan,
  Puerto Rico, May 2016.

\bibitem{mnih2016asynchronous}
V.~Mnih, A.~P. Badia, M.~Mirza, A.~Graves, T.~Lillicrap, T.~Harley, D.~Silver,
  and K.~Kavukcuoglu, ``Asynchronous methods for deep reinforcement learning,''
  in \emph{Proc. Int. Conf. Mach. Learn. (ICML)}, New York, USA, Jun. 2016, pp.
  1928--1937.

\bibitem{schulman2017proximal}
J.~Schulman, F.~Wolski, P.~Dhariwal, A.~Radford, and O.~Klimov, ``Proximal
  policy optimization algorithms,'' \emph{arXiv preprint arXiv:1707.06347},
  2017.

\bibitem{niknam2020federated}
S.~Niknam, H.~S. Dhillon, and J.~H. Reed, ``Federated learning for wireless
  communications: Motivation, opportunities, and challenges,'' \emph{IEEE
  Commun. Mag.}, vol.~58, no.~6, pp. 46--51, Jul. 2020.

\bibitem{yang2020energy}
Z.~Yang, M.~Chen, W.~Saad, C.~S. Hong, and M.~Shikh-Bahaei, ``Energy efficient
  federated learning over wireless communication networks,'' \emph{IEEE Trans.
  Wireless Commun.}, vol.~20, no.~3, pp. 1935--1949, Nov. 2020.

\bibitem{amiri2020federated}
M.~M. Amiri and D.~G{\"u}nd{\"u}z, ``Federated learning over wireless fading
  channels,'' \emph{IEEE Trans. Wireless Commun.}, vol.~19, no.~5, pp.
  3546--3557, Feb. 2020.

\bibitem{li2020harvest}
Y.~Li, R.~Zhao, Y.~Deng, F.~Shu, Z.~Nie, and A.~H. Aghvami,
  ``Harvest-and-opportunistically-relay: Analyses on transmission outage and
  covertness,'' \emph{IEEE Trans. Wireless Commun.}, vol.~19, no.~12, pp.
  7779--7795, Aug. 2020.

\bibitem{stable-baselines3}
A.~Raffin, A.~Hill, A.~Gleave, A.~Kanervisto, M.~Ernestus, and N.~Dormann,
  ``Stable-baselines3: Reliable reinforcement learning implementations,''
  \emph{J. Mach. Learn. Res.}, vol.~22, no. 268, pp. 1--8, Nov. 2021.

\bibitem{schaul2016prioritized}
T.~Schaul, J.~Quan, I.~Antonoglou, and D.~Silver, ``Prioritized experience
  replay,'' in \emph{Proc. Int. Conf. Learn. Represent. (ICLR)}, San Juan,
  Puerto Rico, May 2016, pp. 1--21.

\bibitem{goutay2020deep}
M.~Goutay, F.~A. Aoudia, and J.~Hoydis, ``Deep hypernetwork-based {MIMO}
  detection,'' in \emph{Proc. IEEE Int. Workshop Signal Process. Adv. Wireless
  Commun. (SPAWC)}, Atlanta, USA, May 2020, pp. 1--5.

\end{thebibliography}

\end{document}